\documentclass[12pt,letterpaper,usenames,dvipsnames]{article}
\usepackage{Auxiliary/jheppub}
\usepackage{tcolorbox}
\usepackage{comment}
\usepackage{dsfont}
\usepackage{amsmath,amssymb,amsfonts,amsthm,amscd,setspace}
\usepackage{bm} 
\usepackage{bbm} 
\usepackage{mathbbol} 

\usepackage{graphicx,tikz}
\usepackage[framemethod=TikZ]{mdframed} 
\usepackage{tikz-cd} 
\usetikzlibrary{arrows,snakes,shapes.arrows,decorations.markings}
     \tikzset{>=triangle 90}
     \tikzstyle{gr}=[draw,circle,green!50!black,fill=green!50!black,scale=.6]
     \tikzstyle{Bl}=[draw,circle,blue,scale=.6]
     \tikzstyle{R}=[draw,circle,fill=red,scale=.6]
     \tikzstyle{bl}=[draw,circle,fill=black,scale=.35]
     \tikzstyle{bbc}=[draw,circle,fill=black,scale=.75]
     \tikzstyle{bbcs}=[draw,circle,fill=black,scale=.5]
     \tikzstyle{rc}=[circle,fill=red,scale=.6]
     \tikzstyle{wc}=[draw,circle,scale=.75]

\usepackage{multicol}
\usepackage[export]{adjustbox}
\usepackage{rotating} 
\usepackage{colortbl}
\usepackage{floatrow}

\usepackage{eufrak} 
\usepackage{mathrsfs} 
\usepackage{enumitem}
\usepackage{tcolorbox,array,hhline,arydshln,multirow,floatrow,subfig}

\usepackage{xcolor}
\def\red#1{{\color{red}{#1}}}
\def\blue#1{{\color{blue}{#1}}}
\def\green#1{{\color{black!50!green}{#1}}}

\def\yellow#1{{\color{black!25!yellow!70!red}{#1}}}

\def\nn{\nonumber}

\def\bM{\begin{matrix}}
\def\eM{\end{matrix}}
\newcommand{\bpm}{\begin{pmatrix}}
\newcommand{\epm}{\end{pmatrix}}
\newcommand{\bsm}{\begin{smallmatrix}}
\newcommand{\esm}{\end{smallmatrix}}
\newcommand{\bspm}{\left(\begin{smallmatrix}}
\newcommand{\espm}{\end{smallmatrix}\right)}
\def\bar{\overline}
\def\tilde{\widetilde}
\def\hat{\widehat}

\def\^{\wedge}

\def\so{\mathfrak{so}}
\def\SL{{\rm SL}}

\def\SU{{\rm SU}}

\def\Sp{{\rm Sp}}
\def\sp{\mathfrak{sp}}
\def\USp{{\rm USp}}
\def\U{{\rm U}}
\def\u{\mathfrak{u}}

\def\C{\mathbb{C}}

\def\ff{{\mathfrak f}}

\def\H{\mathbb{H}}

\def\I{\mathbb{I}}

\def\N{\mathbb{N}} 
\def\cN{{\mathcal N}}

\def\cP{{\mathcal P}}
\def\Q{\mathbb{Q}} 

\def\R{\mathbb{R}} 

\def\cS{{\mathcal S}}

\def\cT{{\mathcal T}}

\def\cV{{\mathcal V}}

\def\Z{\mathbb{Z}}
\def\a{{\alpha}}

\def\b{{\beta}}
\def\g{{\gamma}}

\def\D{{\Delta}}

\def\l{{\lambda}}
\def\L{{\Lambda}}

\def\s{{\sigma}}

\def\t{{\tau}}

\def\w{{\omega}}

\def\bar{\overline}
\def\til{\widetilde}
\def\hat{\widehat}

\def\^{\wedge}
\def\I{\mathds{1}}

\def\U{{\rm U}}
\def\SU{{\rm SU}}

\def\SL{{\rm SL}}
 
\def\Sp{{\rm Sp}}

\def\USp{{\rm USp}}

\def\bh{{\boldsymbol h}}

\def\bu{{\boldsymbol u}}

\def\af{\mathfrak{a}}

\def\cf{\mathfrak{c}}

\def\df{\mathfrak{d}}
\def\ef{\mathfrak{e}}
\def\ff{\mathfrak{f}}
\def\gf{\mathfrak{g}}

\def\Sf{\mathfrak{S}}

\def\sof{\mathfrak{so}}
\def\spf{\mathfrak{sp}}
\def\suf{\mathfrak{su}}

\def\uf{\mathfrak{u}}

\def\cN{{\mathcal N}}

\def\cP{{\mathcal P}}

\def\cS{{\mathcal S}}

\def\cT{{\mathcal T}}

\def\cV{{\mathcal V}}

\def\Msc{\mathscr{M}}

\def\C{\mathbb{C}}

\def\H{\mathbb{H}}
\def\N{\mathbb{N}}

\def\Q{\mathbb{Q}}  
\def\R{\mathbb{R}} 
 
\def\Z{\mathbb{Z}}

\def\beq{\begin{equation}}
\def\eeq{\end{equation}}
\def\nn{\nonumber}
\newcommand{\bpmat}{\begin{pmatrix}}
\newcommand{\epmat}{\end{pmatrix}}
\newcommand{\bsmat}{\begin{smallmatrix}}
\newcommand{\esmat}{\end{smallmatrix}}
\newfloatcommand{capbtabbox}{table}[][\FBwidth]

\def\ccg{\cellcolor{green!07}}
\def\ccgy{\cellcolor{black!25!green!20!yellow!10}}

\def\rcg{\rowcolor{green!07}}
\def\rcy{\rowcolor{black!25!yellow!10}}

\def\rcr{\rowcolor{red!07}}

\definecolor{amaranth}{rgb}{0.9, 0.17, 0.31}
\definecolor{coolblack}{rgb}{0.0, 0.18, 0.39}
\definecolor{gold(web)(golden)}{rgb}{1.0, 0.84, 0.0}
\definecolor{deepcarmine}{rgb}{0.66, 0.13, 0.24}

\title{The rank 2 classification problem II:\\
mapping scale-invariant solutions to SCFTs}
\author[1]{Philip C. Argyres}
\author[2]{Mario Martone}
\affiliation[1]{University of Cincinnati, Physics Department, PO Box 210011, Cincinnati OH 45221}
\affiliation[2]{Dept.\ of Mathematics, King's College London, The Strand, London WC2R 2LS, UK}
\emailAdd{philip.argyres@gmail.com}
\emailAdd{mario.martone@kcl.ac.uk}

\abstract{This is the second of a series of papers outlining an approach to the classification of $\cN{=}2$ superconformal field theories at rank 2 via a systematic analysis of their Coulomb branches, mathematically described by special K\"ahler scale invariant geometries.  Here we describe how to make the translation between geometry and field theory data.  We apply this strategy to the special K\"ahler geometries found in \cite{Argyres:2022lah} where we made strong simplifying assumptions on the form of the solutions. Remarkably, we find that our bottom-up classification strategy pays off even in this simplified setup.  All scale invariant solutions in \cite{Argyres:2022lah} have an interpretation as the Coulomb branch of at least one $\cN{=}2$ superconformal field theory, many matching what is already known.  But we also predict the existence of three new rank 2 theories, and our methods improves our understanding of the moduli space structure of three more.  We characterise and discuss in detail each of these cases. To our knowledge none of the new theories have known string theoretic or higher dimensional realisations.}

\begin{document}
\maketitle 

\section{Introduction}

The general program of classifying $\cN{=}2$ superconformal field theories (SCFTs) from moduli space constraints organises theories based on the dimension, aka \emph{rank}, of their Coulomb branches (CBs).
The idea is that the complexity of the space of SCFTs increases as the rank increases. 
One uses the fact that CBs of $\cN{=}2$ SCFTs are constrained by $\cN{=}2$ superconformal invariance to be scale invariant special K\"ahler (SISK) varieties \cite{Seiberg:1994I, Seiberg:1994II, Freed:1997dp} which can then be classified.
Such a classification then allows one to map out the space of $\cN{=}2$ SCFTs.
It is this last step, which involves deducing SCFT properties from SISK geometries, that is the focus of this paper.

We will show that all the solutions obtained in \cite{Argyres:2022lah} have the interpretation as the CB of at least one $\cN{=}2$ SCFT, and most have multiple such interpretations.%
\footnote{While all CBs of $\cN{=}2$ SCFTs are SISK varieties, it can happen that a SISK variety has no interpretation as the CB of any $\cN{=}2$ SCFT.  While a few such examples are known, they seem to be rare.}
This well-known ambiguity in the relation between SISK geometries and SCFTs \cite{Seiberg:1994I, Seiberg:1994II} is lifted upon considering the ways the scale invariant geometries can be deformed while preserving a special K\"ahler (SK) structure and the asymptotic geometry at metric infinity.
These geometric deformations correspond to turning on relevant deformations of the SCFTs such as mass parameters. 
Thus the task at hand is roughly divided into two steps,
\begin{itemize}
    \item[$a$)] the classification of SISK geometries of a given dimension $r$, and    
    \item[$b$)] the analysis of their SK deformations.
\end{itemize}

For one dimensional (a.k.a.\ rank 1) CBs, both $a)$ and $b)$ have been solved. 
In this case there are seven SISK geometries which are a subset of the Kodaira classification of degeneration of elliptic curves \cite{KodairaI, KodairaII}, the elliptic curve arising as part of the description of the SK geometry. 
Carrying out $b)$ even in this simple case turned out to be a very non-trivial task which was completed only a few years ago \cite{Argyres:rank1I, Argyres:rank1II, Argyres:rank1III, Argyres:rank1IV}, and gave 27 distinct geometries upon deformation, all but two of which correspond to SCFTs. 
This large number of geometries at rank 1 was unexpected, and it in fact predicted the existence of many new theories \cite{Garcia-Etxebarria:2015wns, Aharony:2016kai, Chacaltana:2015bna, Ohmori:2018ona, Apruzzi:2020pmv}. 
While considerably richer than expected, the rank 1 case is very special: all the rank 1 SISK geometries are \emph{isotrivial}, \emph{i.e.}, are flat cones.
By contrast, at rank 2 isotrivial geometries are expected to be a small subset of the allowed scale invariant geometries. 
Hoping to extrapolate the geography of higher rank SCFTs from those at rank 1 is in many ways as hopeful as trying to extrapolate the classification of classical Lie algebras from knowledge of the rank 1 algebra. 

In a short series of three papers including this one and \cite{Argyres:2022lah, Argyres:2022fwy}, we set up this program in the next simplest case, rank 2.  
At rank 2, step $a)$ is a clearly delineated mathematical problem, which, while considerably more complex than at rank 1, still has many special simplifying features (analogous to low-rank exceptional isomorphisms of classical Lie algebras).
This is explained extensively in \cite{Argyres:2022lah} in which we work out the complete list of scale-invariant solutions for what we call the \emph{polynomial ansatz}, which we reproduce in table \ref{tab:solsr1} below, and which is expected to capture only a relatively small subset of the rank 2 SISK geometries. 
The approach of \cite{Argyres:2022lah} was inspired by a previous attempt by one of the authors at tackling this problem \cite{Argyres:2005x6, Argyres:2005x5}.

In this paper we instead focus on step $b)$, the systematic analysis of possible mass deformations of each of these solutions. 
For step $b)$ at rank 2 it is a challenge to perform any direct calculations of deformations preserving an SK structure, let alone classify them.
To accomplish this task we sidestep direct calculation of deformations by relying heavily on our improved understanding of the stratification structure of the singular locus of higher dimensional CBs \cite{Martone:2020nsy, Argyres:2020wmq} together with insights from the rank 1 analysis \cite{Argyres:rank1I, Argyres:rank1II, Argyres:rank1III, Argyres:rank1IV}. 

More specifically, \cite{Martone:2020nsy, Argyres:2020wmq} developed a calculus --- reviewed below ---  which enables a translation between the geometric data of the SISK solution to the physical data characterising each $\cN{=}2$ SCFT, \emph{e.g.} central charges, flavour symmetries, etc. 
This map is not one to one; rather there are multiple SCFT data assignments consistent with each SISK variety. 
In this paper we make this statement sharp, identifying precisely where the ambiguity lies, and are able to identify a new consistency requirement which picks out the subset of \emph{physical} SCFT data assignments. 
We then make a comparison between the results we find and the current status of knowledge of rank 2 $\cN{=}2$ SCFTs, mapped out recently in \cite{Martone:2021r2}. 
Largely we reproduce what is already known but we also find three new SCFTs, and improve our understanding of three more; we describe each case in detail.

A major advantage of our CB based formalism is that the CB is only deformed but not lifted by relevant deformations \cite{Green:2010da, Argyres:rank1I, Cordova:2016defs} thus we are able to also map out renormalization group (RG) flows among various different theories. 
This worked spectacularly well at rank 1 where we found that all known theories are organised in four sets which are connected by RG flows and which we call \emph{series}. 
Each series in rank 1 is labeled by a CB invariant, namely, ``$I_1$", ``$I_4$", ``$I_1^*$", and ``$IV^*$", which identifies the largest singularity which is left undeformed by the RG flow characterising the given series \cite{Argyres:rank1I, Argyres:rank1II, Argyres:rank1III, Argyres:rank1IV}.  
Our analysis here allows us to make a surprising connection with this rank 1 picture: we find that to a good extent also rank 2 theories can be organised by similarly labeled series. 
One difference in the rank 2 case is that we find two $I_4$ series, which we label $I_4$ and $\widetilde{I_4}$, which from the rank 1 point of view are distinguished by a subtle freedom in normalising the beta function of IR-free $\SU(2)$ gauge theories which is explained in \cite[Section 4.2]{Argyres:rank1I} and briefly reviewed in appendix \ref{appB}. Here we find new entries for the rank 2 $\tilde{I}_4$ and $I_1^*$ series while the $IV^*$ series is new, though the latter currently consists of only one SCFT. 
The results are summarised in table \ref{tab:knownSWr2}.

This surprising result suggests that the somewhat involved notion of deformation patterns of rank 1 theories developed in \cite{Argyres:rank1I, Argyres:rank1II, Argyres:rank1III, Argyres:rank1IV} is the right physical notion to organise SCFTs also in the higher rank case --- where now the rank 1 theory appears on the components of the singular locus. 
This pattern organizing the rank 2 RG series can be partially seen from the point of view of string or brane constructions as well. 
Recall that the four rank 1 RG series were later understood in terms of certain generalised orbifold constructions in F theory called \emph{S-folds} \cite{Aharony:2015oyb, Apruzzi:2020pmv, Giacomelli:2020jel, Heckman:2020svr, Giacomelli:2020gee, Bourget:2020mez}. 
This S-fold picture is simpler than the original analysis of mass deformations and provides a considerably more homogeneous framework for understanding the four series. 
Unfortunately the correspondence between S-folds and $I_1$--$IV^*$ series does not seem to hold beyond rank 1. 
Many of the rank 2 S-folds are of the split type discussed in \cite{Argyres:2022lah}, and in fact have characteristic dimension \cite{Cecotti:2021ouq} $\varkappa\neq\{1,2\}$, which is not the case in the RG series we discuss here. A more promising route for mapping out the structure we find, is the recently introduced \emph{SW}-folds \cite{Heckman:2022suy}  a slight generalisation of the \emph{S}-fold associated to some sort of orbifolds by a cyclic group, specifically $I_1\leftrightarrow\I$, $I_4\leftrightarrow\Z_2$ (both for the $I_4$ and $\tilde{I}_4$, see table \ref{tab:knownSWr2}), and $I^*_1\leftrightarrow\Z_3$.\footnote{We thank Craig Lawrie for clarifying this interpretation to us.} 
Our analysis suggests that also the last series is associated to such orbifold structure: $IV^*\leftrightarrow\Z_4$, though this has no analog on the string theory side. In fact none of the new SCFTs that we find, which importantly cannot be reached by mass deformations from currently known SCFTs --- see figure \ref{MapR2} --- and include the lone SCFT in the $IV^*$ series, have no known string theoretic realisation.

The paper is organised as follows. 
Section \ref{sec:2} reviews the main results from the CB stratification analysis with particular attention to the task of translating from the data accessible from the SISK results reported in table \ref{tab:solsr1} to SCFT data. 
Section \ref{sec:3} discusses in detail the deformations of the solutions found in \cite{Argyres:2022lah} which lead to new rank 2 SCFTs. The remaining cases, which reproduce remarkably well the current knowledge of rank 2 $\cN{=}2$ SCFTs, are reported in appendix \ref{appA}. 
Section \ref{sec5} concludes with a very brief discussion, largely summarized in figure \ref{MapR2}, updating the map of rank 2 $\cN{=}2$ SCFTs and their RG flows.
Finally, appendix \ref{appB} reviews and collects results about rank 1 SCFTs and IR-free gauge theories which are used throughout the paper.

\begin{table}[t!]
\begin{adjustbox}{center,max width=1\textwidth}
$\def\arraystretch{1.0}
\begin{array}{c|c:c:c|c}
\multicolumn{5}{c}{\Large \textsc{CBs given by polynomial families of genus 2 RSs}}\\
\hline
\hline
 \D_{u,v}&\ \red{\mathfrak{S}}_{\rm knot} \ \, &\quad \red{\mathfrak{S}}_u \quad\, &\quad \red{\mathfrak{S}}_v \quad\,
&\textrm{Form of the curve in canonical frame}
\\
\cdashline{1-5}

 \{6,8\}
&I_1&\varnothing & I_6^*
&\ccg y^2=x\, (u\, x-v)\,(x^4+u\, x-v)\\[1mm]

\{4,10\}
&I_1&\varnothing & II^*
&y^2=x^5+ (u\, x-v)^3\\[1mm]

\{4,6\}
&I_1&I_2 & I_4^*
&\ccg y^2=x\, (u\,x-v)\, (v\, x^3 +u\, x-v)\\[1mm]
 \{4,5\}

&I_1&\varnothing & I_{10}
&\ccg y^2=(u\, x-v)(x^5+u\, x-v) \\[1mm]

 \{3,5\}

&I_1&\varnothing & I_3^*
&y^2=x(x^5+(u\, x-v)^2) \\[1mm]
 \{3,4\}

&I_1& I_2 & I_8
&\ccg y^2=(u\, x-v)(x^4+u\, x-v)\\[1mm]

\rcy \{2,4\}

&(I_1)^2&\varnothing & I_2^*
& y^2=x(x^4+\t\,x^2\,(u\, x-v)+(u\,x-v)^2)\\[1mm]

\rcy \{2,3\} 

&(I_1)^2&\varnothing & I_6
&y^2=x^6+\t x^3\, (u\,x-v)+ (u\, x-v)^2\\[1mm]

\rcy \{2,2\}

&(I_2)^3&I_2 & I_2
& \ccgy y^2=(u\, x-v)(x^5 + \t_1 x^3 + \t_2 x^2 + \t_3 x + \t_4)\\[1mm]

\{\frac32,\frac52\} 

&I_1&\varnothing & I_5
&y^2=x^5+(u\,x-v)^2\\[1mm]

\{\frac43,\frac53\}

&I_1&\varnothing & I_2 
&y^2=x(x^5+u\,x-v)\\[1mm]

\{\frac65,\frac85\}

&I_1&\varnothing & I_2
&y^2=x(x^4+u\,x-v)\\[1mm]

\{\frac54,\frac32\} 
&I_1&\varnothing & \varnothing
&y^2=x^6+u\,x-v\\[1mm]

\{\frac87,\frac{10}7\} 

&I_1&\varnothing & \varnothing
&y^2=x^5+u\,x-v\\[.5mm]

\hline
\hline

\end{array}$
\caption{\label{tab:solsr1}
All rank 2 scale invariant CB geometries described by genus 2 SW curves with polynomial coefficients. 
The first column gives the CB scaling dimensions.
The next three columns list the monodromies around each co-dimension one stratum. 
$\varnothing$ means that the corresponding stratum is not actually part of the singular locus and the power means that there are multiple components of that given form. 
The last column gives the binary-sextic plane curve; in green are the solutions which are new to this paper, and in yellow the solutions which correspond to Lagrangian theories.
}
\end{adjustbox}
\end{table}

\section{From CB to SCFT data}
\label{sec:2}

The key tool to be able to translate from geometric to physical data is the notion of stratification of the singular locus of the CB. 
We start by reviewing the basics and then explain how to leverage the structure of the stratification to obtain the SCFT data compatible with it. 
Throughout this work we will assume that the CB has no complex singularities, it is thus topologically $\C^r$, and it is principally polarised, see \cite{Argyres:2022lah} or \cite{Argyres:2022fwy} for a critical assessment of these assumptions. 

\subsection{Stratification of the Coulomb branch}

The low-energy effective theory on a generic point of the CB is an $\cN{=}2$ $\U(1)^r$ gauge theory. In the case in which the theory is superconformal, the $\U(1)_r$ and the $\R^+$ scaling symmetry combine to give a holomorphic $\C^*$ action on the CB, which we will often call the \emph{scaling action}, and under which all the structures of the CB of an SCFT are equivariant. 
In particular it is always possible to choose coordinates $(u_1,...,u_r)$ of the CB which diagonalize the scaling action,
\beq
\C^*: (u_1,...,u_r) \mapsto (\l^{\D_{u_1}}u_1,...,\l^{\D_{u_r}}u_r),\quad 
\l\in\C^* \quad \text{and} \quad 0<\D_{u_i}\in\Q ,
\eeq
where the $\D_{u_i}$ are the coordinates' scaling dimensions. 
We will always assume that we have chosen such a set as our coordinates.

\begin{table}[t!]
\begin{adjustbox}{center,max width=.85\textwidth}
$\def\arraystretch{1.0}
\begin{array}{c:cc|c:cc|c:c:c:c}
\multicolumn{10}{c}{\Large \textsc{Rank 2 theories with $\varkappa=$1 or 2 and known SW curve}}\\
\hline
\hline
\multicolumn{3}{c|}{ \text{Moduli Space}} &
\multicolumn{3}{c|}{ \text{Flavor and central charges}} 
\\[1mm]
 \D_{u,v}&\ \ d_{\text{HB}}\ \ &\ \  h\ \  
&\quad \ff\quad &\ \ 24a\ \ & 12c 
&\multicolumn{4}{c}{\multirow{-2}{*}{\text{Curve information}}}
\\[1.5mm]
\cdashline{1-10}

\multicolumn{10}{c}{\ef_8-\sof(20)\ -\ I_1\ \text{series} }\\
\hline

 \{6,8\}
& 46 & 0
&\sof(20)_{16} & 202 & 124  
&\multicolumn{4}{c}{  y^2=x\, (u\, x-v)\,(x^4+u\, x-v)}\\
 \{4,10\}
 & 46 & 0
&[\ef_8]_{20} & 202 & 124 
&\multicolumn{4}{c}{y^2=x^5+ (u\, x-v)^3}\\
\{4,6\}
&30 & 0
&\suf(2)_8 \times \sof(16)_{12} & 138 & 84 
&\multicolumn{4}{c}{ y^2=x\, (u\,x-v)\, (x^3+u\, x-v)}\\
 \{4,5\}
&  26 & 0
&\suf(10)_{10} & 122  & 74 
&\multicolumn{4}{c}{  y^2=(u\, x-v)(x^5+u\, x-v)} \\

 \{3,5\}
&  22 & 0
&\sof(14)_{10}\times \uf(1) & 106  & 64 
&\multicolumn{4}{c}{y^2=x(x^5+(u\, x-v)^2) } \\
 \{3,4\}
&18 & 0
&\suf(2)_6\times \suf(8)_8 & 90  & 54  
&\multicolumn{4}{c}{  y^2=(u\, x-v)(x^4+u\, x-v)}\\
\rcy \{2,4\}
& 14 & 0 
& \sof(12)_8 & 74 & 44 
&\multicolumn{4}{c}{y^2=x(x^4+\t\,x^2\,(u\, x-v)+(u\,x-v)^2)}\\
\rcy  \{2,4\} 
& 11 & 1 
& \sof(8)_8\times\suf(2)_5 & 75 & 42  
&\multicolumn{4}{c}{{\rm Degenerate\ genus\ 2\ RS}}\\
\rcy \{2,3\} 
& 10 & 0
& \uf(6)_6 & 58 & 34
&\multicolumn{4}{c}{y^2=x^6+\t x^3\, (u\,x-v)+ (u\, x-v)^2}\\
\rcy \{2,2\}
&  6 & 0
& \suf(2)_4^5 & 42 & 24
&\multicolumn{4}{c}{\ \, y^2=(u\, x-v)(x^5 + \t_1 x^3 + \t_2 x^2 + \t_3 x + \t_4)\ \,}\\

\{\frac32,\frac52\} 
& 6 & 0  
& \suf(5)_5 & 42 & 24 &
\multicolumn{4}{c}{y^2=x^5+(u\,x-v)^2}\\

 \{\frac43,\frac53\}
& 2 & 0
&\suf(2)_{\frac{10}3} \times \uf(1) & 26 & 14 &
\multicolumn{4}{c}{y^2=x(x^5+u\,x-v)}\\

 \{\frac65,\frac85\}
& 1 & 0 &
\suf(2)_{\frac{16}5} & \frac{114}5 & 12 &
\multicolumn{4}{c}{y^2=x(x^4+u\,x-v)}\\
 \{\frac54,\frac32\} 
& 1 & 0
& \uf(1) & 22 & \frac{23}2 &
\multicolumn{4}{c}{y^2=x^6+u\,x-v}\\
 \{\frac87,\frac{10}7\} 
& 0 &0  & \varnothing
& \frac{134}7 & \frac{68}7 &
\multicolumn{4}{c}{y^2=x^5+u\,x-v}\\[.5mm]
\hline\hline


\multicolumn{10}{c}{\suf(6)\ -\ I_4\  \text{series}}\\
\hline
 \{6,8\} 
 & 23 &1  &
\suf(6)_{16}{\times}\suf(2)_9 & 179 & 101 
&\multicolumn{4}{c}{  y^2=x\, (u\, x-v)\,(x^4+u\, x-v)}\\
 \{4,6\}
& 13 & 1 &
\suf(4)_{12}{\times} \suf(2)_7{\times}\uf(1) & 121 & 67 
&\multicolumn{4}{c}{  y^2=x\, (u\,x-v)\, (x^3 + u\, x-v)}\\
 \{4,5\}
& 11 & 0 &
\suf(3)_{10}{\times} \suf(3)_{10}{\times}\uf(1) & 107 & 59  
&\multicolumn{4}{c}{  y^2=(u\, x-v)(x^5+u\, x-v)}\\
 \{3,5\}
& 8 & 1 &
\suf(3)_{10}{\times} \suf(2)_6{\times}\uf(1) & 92 & 50  
&\multicolumn{4}{c}{y^2=x(x^5+(u\, x-v)^2) }\\
 \{3,4\}
& 6 & 0 &
\suf(2)_8{\times} \suf(2)_8{\times}\uf(1)^2 & 78 & 42 
&\multicolumn{4}{c}{  y^2=(u\, x-v)(x^4+u\, x-v)}\\
\rcy \{2,3\} 
& 2 & 0
& \uf(1)\times \uf(1) & 49 & 25  
&\multicolumn{4}{c}{y^2=x^6+\t x^3\, (u\,x-v)+ (u\, x-v)^2}\\[.5mm]

\hline\hline


\multicolumn{10}{c}{\spf(14)\ - \tilde{I_4}\ \text{series}}\\
\hline
 \{6,8\}
&  29 & 7
&\spf(14)_9 & 185 & 107 
&\multicolumn{4}{c}{  y^2=x\, (u\, x-v)\,(x^4+u\, x-v)}\\
\rcg \{4,10\}
 & 21 & 5
&\red{\spf(10)}_{11} & 177 & 99
&\multicolumn{4}{c}{y^2=x^5+ (u\, x-v)^3}\\

 \{4,6\}
& 17 & 5
&\suf(2)_8 \times\spf(10)_7& 125  & 71 
&\multicolumn{4}{c}{  y^2=x\, (u\,x-v)\, (x^3 +u\, x-v)}
\\

 \{3,5\}
& 11 & 4
&\spf(8)_6\times \uf(1) & 95  & 53 
&\multicolumn{4}{c}{y^2=x(x^5+(u\, x-v)^2) }
\\
\rcy \{2,4\}
& 6 & 3
& \spf(6)_5 & 65 & 35 
&\multicolumn{4}{c}{y^2=x(x^4+\t\,x^2\,(u\, x-v)+(u\,x-v)^2)}\\

\hline\hline


\multicolumn{10}{c}{\suf(5)\ - I_1^*\ \text{series}}\\
\hline
 \{6,8\}
&  14 & 0
&\suf(5)_{16} & 170 & 92 
&\multicolumn{4}{c}{  y^2=x\, (u\, x-v)\,(x^4+u\, x-v)}
\\
\rcg \{4,10\}
 & 10 & 4
& \red{\suf(4)}_{11}& 166 & 88
&\multicolumn{4}{c}{y^2=x^5+ (u\, x-v)^3}\\
 \{4,6\}
& 6 & 0
&\suf(3)_{12}\times \uf(1)& 114  & 60 
&\multicolumn{4}{c}{  y^2=x\, (u\,x-v)\, (x^3 +u\, x-v)}
\\

 \{3,5\}
& 3 & 0
& \suf(2)_{10}\times\uf(1) & 86  & 44 
&\multicolumn{4}{c}{y^2=x(x^5+(u\, x-v)^2) }
\\

\hline\hline


\multicolumn{10}{c}{ \suf(3)\ - IV^*\ \text{series}}\\
\hline
\rcg \{4,10\}
 & 4 & 3
&\red{\suf(3)}_{11} & 160 & 82
&\multicolumn{4}{c}{y^2=x^5+ (u\, x-v)^3}

\\
[.5mm]
\hline
\hline

\end{array}$
\caption{\label{tab:knownSWr2}
Updated list of rank-2 theories, the ones shaded in \green{green} are, to our knowledge, new. The first 6 columns give the SCFT data while the last column gives the SW curve. We shade in \yellow{yellow} $\cN=2$ Lagrangian theories.
}
\end{adjustbox}
\end{table}

The CB is a singular SK variety where the singular locus $\cV$ has complex co-dimension one. 
The singular locus is physically interpreted as the locus where extra states which carry electric or magnetic charges under $\U(1)^r$ become massless. 
It is a stratified space \cite{Argyres:2020wmq} which has a lot of structure playing an important role in our analysis. 
Specifically:
\begin{itemize}
    \item It has in general multiple connected components, each algebraic and characterised by a different set of charged states becoming massless. Scale invariance further imposes that each component has to be closed under the scaling action. This in particular implies that the algebraic form of each connected component is,
    \beq\label{PolComp}
    \cP(u_1,...,u_r)=0 ,
    \eeq
    where $\cP$ is an irreducible homogeneous polynomial in the CB coordinates $(u_1,...,u_r)$.
    
    \item It contains components of increasingly higher complex co-dimension, the superconformal vacuum, \emph{i.e.}, the origin of the CB, being the sole $r$ complex co-dimensional locus.
    
    \item The low energy effective theory on each of component of complex co-dimension $\ell$ is richer than in the generic locus and it is described by a rank $\ell$ conformal or IR-free field theory.
\end{itemize}

As we will describe in detail in the next section, the translation between geometric and physical data can be done largely by only studying the complex co-dimension one locus. 
This then entails identifying the number of connected components indexed by a set $I$ together with their associated rank 1 theories,
\begin{align}
    \cV_i = \{ \cP_i(\bu)=0 \}, \quad i\in I 
    \qquad \Leftrightarrow \qquad \cT^i_{\rm rank-1} .
\end{align}
Here $\cP_i$ is homogeneous as in \eqref{PolComp}, and the  rank 1 theory $\cT^i_{\rm rank-1}$ ---  which can be either IR-free or an SCFT ---  describes the extra states becoming massless at $\cV_i$.

More specifically, to characterise the physical data we only need the scaling dimension of the polynomial
\beq
\D_i^{\rm sing} \doteq \D\big[\cP_i(\bu)\big] ,
\eeq
and the $b$ factor of the rank 1 theory \cite{Martone:2020nsy} which are also tabulated in table \ref{tab:r1theories},
\beq\label{bir1}
b^i_{\rm rank-1}:=\frac{12 c_i-2 - h_i}{\D_u^i} .
\eeq
Here $c_i$, $h_i$ and $\D_u^i$ are, respectively, the $c$ central charge, the quaternionic dimension of the extended CB \cite{Argyres:rank1IV} and the CB scaling dimension of $\cT_{\rm rank-1}^i$. 
To make the discussion less abstract we will soon apply this formalism to a specific example. 
But before we do that, let's specify our discussion to the case of rank 2 where this structure becomes considerably simpler.

\subsection{Connected components and rank 1 theories at rank 2}

Perhaps the most simplifying feature of rank 2 is that there are only a few possible forms of the $\cP_i(\bu)$ that we need to consider since the set of homogeneous irreducible polynomial in two complex variables are extremely constrained. 
In fact, specifying our choice of CB coordinates to rank 2,
\beq
(u_1,u_2) \doteq (u,v),\qquad \D_u\leq \D_v ,
\eeq
there are only three such polynomials,
\begin{align}\label{unkno}
    {\rm unknotted}:&\quad u=0\quad {\rm or}\quad v=0\\
    {\rm knotted}:&\quad u^p+\a\, v^q=0\quad \left\{
    \begin{array}{l}
        p/q=\D_v/\D_u\quad {\rm and}\\\label{kno}
        {\rm gcd}(p,q)=1  
    \end{array}
    \right.
\end{align}
where $\a\in \C^*$.
We will use the names \emph{knotted} and \emph{unknotted} to refer to a co-dimension one singular component described by \eqref{unkno} and \eqref{kno}, respectively \cite{Argyres:2018Cody}. 
It then follows that
\beq\label{Deltasing}
\D^{\rm sing}_i=
\begin{cases}
\D_u \ \text{or}\ \D_v , & \text{unknotted}, \\
p\D_u=q\D_v , & \text{knotted} .
\end{cases}
\eeq

Another simplifying feature follows from the form of the rank 2 SISK solutions. As described in \cite{Argyres:2022lah,Argyres:2022fwy}, they can always be put in the \emph{hyperelliptic form}:
\beq\label{hyper}
y^2=c(x,u,v)
\eeq
where $c(x,u,v)$ is a sixth or fifth order polynomial in $x$ with coefficients meromorphic in $(u,v)$. 
Then $\cV$ is given by the zeros of the $x$ discriminant of the right hand side of \eqref{hyper} which factorises as the product of knotted and unknotted components, each representing a different connected component,
\beq\label{DiscGen}
{\rm Disc}_x[c(x,u,v)]=u^{s_u}v^{s_v}\prod_i (u^p+\a_i\, v^q)^{s_i} .
\eeq
Here the index $i$ runs over the knotted components of the singular locus $\cV$. 
In general the order of vanishing of the discriminant at each component, \emph{i.e.}, the tuple $(s_u,s_v,s_i)$, can be greater than one and can be used to identify the rank 1 theory $\cT_{\rm rank-1}^i$ describing the massless states along the given component. 
We now explain how this works.

The geometric datum which constrains the rank 1 theory $\cT_{\rm rank-1}^i$ supported on $\cV_i$ is the monodromy of the special coordinates along a small loop $\g_i$ which links only $\cV_i$ once. This monodromy, $\Msc_{\g_i}\in\Sp(4,\Z)$, can always be brought by a choice of EM duality basis to a form where along the diagonal it decomposes into an identity $2\times 2$ matrix and an element of $\SL(2,\Z)$ (in the rank $r$ case the identity matrix becomes $2(r{-}1)\times 2(r{-}1)$) \cite{Argyres:2018Cody},
\beq
\Msc_{\g_i}=\left(
\begin{array}{c|c}
\Msc^i_{SL(2,\Z)}     & * \\
\hline
*     &  \I_{2\times 2}
\end{array}
\right) .
\eeq
The precise form of the off-diagonal blocks is inessential. Henceforth we will call the $\Msc^i_{SL(2,\Z)}$ piece the \emph{linking monodromy} of the component linked by the curve $\g_i$. Since the CB of the rank 1 theory $\cT_{\rm rank-1}^i$ is identified with the complex plane transverse to $\cV_i$, $\Msc^i_{SL(2,\Z)}$ is precisely the monodromy which characterises its singular behaviour. 

More specifically, $\Msc^i_{SL(2,\Z)}$ only specifies the scale invariant limit of the rank 1 CB. 
In this sense, it does not give any information about the mass deformation of the geometry and for a given $\Msc^i_{SL(2,\Z)}$ there are multiple $\cT^i_{\rm rank-1}$ assignments compatible with it. 
For example, the $I_0^*$ Kodaira type corresponds to the scale invariant geometry for a $\cN{=}2$ $\SU(2)$ gauge theory with four hypermultiplets in the fundamental as well as for a single hypermultiplet in the adjoint --- $\cT^{(1)}_{D_4,1}$ and $\blue{\cS^{(1)}_{\varnothing,2}}$ in the notation of table \ref{tab:r1theories}. 
Or the $III^*$ Kodaira type corresponds to the scale invariant CB geometry of three non-Lagrangian $\cN{=}2$ SCFTs: $\cT^{(1)}_{E_7,1}$, $\cS^{(1)}_{D_4,2}$ and $\green{\cS^{(1)}_{\varnothing,3}}$; see appendix \ref{appB} for more details. The ambiguity between scale invariant geometries and SCFT data arises precisely from these multiple possible interpretations of the scale invariant limit of rank 1 geometries.

\begin{table}[t!]
\centering
$\begin{array}{|c|l|c|c|c|}
\hline
\multicolumn{5}{|c|}{\text{\bf Scaling behaviors near rank 1 singularities}}\\
\hline\hline
\text{Name} & \multicolumn{1}{c|}{\text{planar SW curve}} & \ \text{ord}_0(D^\L_{x})\ \ &\ \D(u)\ \ &M_0\\
\hline
II^*   &\parbox[b][0.45cm]{4cm}{$\ y^2=x^3+u^5$}             
&10 &6 &ST \\
III^*  &\ y^2=x^3+u^3x &9 &4&S \\
IV^*  &\ y^2=x^3+u^4 &8 &3 &(ST)^2\\
I_0^* &\ y^2=\prod_{i=1}^3\left(x-e_i(\t)\, u\right)
&6 &2 &-\I\\
IV &\ y^2=x^3+u^2 &4 &3/2 & (ST)^{-2} \\
III &\ y^2=x^3+u x &3 &4/3 & S^{-1} \\
II  &\ y^2=x^3+u &2 &6/5 &  (ST)^{-1} \\
\hline
\hline
I^*_n\ \ (n{>}0) &
\parbox[b][0.45cm]{5cm}{
$\ y^2=x^3+ux^2+\L^{-2n}u^{n+3}\ \ $}
& n+6 & 2 &-T^n\\
I_n\ \ (n{>}0)    &\ y^2=(x-1)(x^2+\L^{-n}u^n)  
& n     & 1&T^n \\[0.5mm]
\hline
\end{array}$
\caption{\label{tab:Kodaira} Scaling forms of rank 1 planar special Kahler singularities, labeled by their Kodaira type (column 1), a representative family of elliptic curves with singularity at $u=0$ (column 2), order of vanishing of the discriminant of the curve at $u=0$ (column 3), mass dimension of $u$ (column 4), a representative of the $\SL(2,\Z)$ conjugacy class of the monodromy around $u=0$ (column 5) where $S$ and $T$ are the usual generators of $\SL(2,\Z)$.  The first seven rows are scale invariant.  The last two rows give infinite series of singularities which have a further dimensionful parameter $\L$ so are not scale invariant; they can be interpreted as IR free theories.}
\end{table}

In rank 1, the order of vanishing of the discriminant can be defined unambiguously and it directly gives information about the corresponding $\Msc^i_{\SL(2,\Z)}$; see table \ref{tab:Kodaira}.
We notice that in many cases also the order of vanishing of the discriminant of the rank 2 curve at the component $\cV_i$ provides similar information about $\Msc^i_{\SL(2,\Z)}$. 
For example if we had a curve $\tilde{c}(x,u,v)$ with discriminant:
\beq
{\rm Disc}_x[\tilde{c}(x,u,v)]=u^8 (u^p+v^q)
\eeq
we would read off that the unknotted component $u=0$ has order of vanishing eight and from table \ref{tab:Kodaira} we would infer that $\Msc^{u=0}_{SL(2,\Z)}=\{T^8,-T^2,(ST)^2\}$, that is the rank 1 CB singularity is of Kodaira type $I_8$, $I_2^*$, or $IV^*$. 
Notice that the knotted component is instead unambiguous since the only linking monodromy compatible with an order of vanishing equal to one is $T$, corresponding to Kodaira type $I_1$. 
We lift the ambiguity among different possibilities simply by directly computing the linking monodromy from the algebraic form of the curve. 
This is something we will resort to often below. 

Summarising, a simple analysis of the SW curve in hyperelliptic form gives the set of components, $\cV_i$, of the singular locus, and for each component also the following information:
\beq\label{compo}
\cV_i:\quad (\cP_i(\bu),s_i,\Msc_{SL(2,\Z)}^i) .
\eeq
But as stated above, and will be reviewed even more carefully below, it is the datum of the rank 1 theory $\cT^i_{\rm rank-1}$ which determines the SCFT data. 
Thus an extra step is needed, enumerating all $\cT^i_{\rm rank-1}$ compatible with \eqref{compo} for all the components of the singular locus. 
We will denote these assignments for all $i$ with curly brackets: $\{\cT_{\rm rank-1}^i\}$ and call each choice consistent with the linking monodromies of the curve, a \emph{deformation} of the scale invariant curve.  

Determining the consistent deformations requires one to be well-versed in rank-1-ology.
If the linking monodromy is of finite order (the first seven entries in table \ref{tab:Kodaira}), so the plane transverse to the singular component is a scale invariant rank 1 CB, then to find the compatible theories it is enough to carefully analyse the SCFTs in \cite{Argyres:rank1IV} compatible with that singularity (see \emph{e.g.} table \ref{tab:r1theories}). 
For IR-free theories, corresponding to $I_n$ or $I^*_n$ singularities with $n>0$, the procedure is more involved and requires listing all possible matter assignments compatible with:
\begin{itemize}
    \item {\bf Dirac Quantization:} The charges of matter fields have to be all mutually commensurate. How this rule applies to the physical interpretation of rank 1 singularities has been studied in depth in \cite{Argyres:rank1I, Argyres:rank1II, Argyres:rank1III, Argyres:rank1IV}.
    \item {\bf Beta function:} The $I_n$ and $I_n^*$ singularities should be interpreted, respectively, as $\U(1)$ and $\SU(2)$ theories with beta function equal to $n$. The allowed freedom is reviewed in appendix \ref{appB}.
\end{itemize}

\subsubsection{An example in detail: $\D_{u,v}=\{4,5\}$}

We illustrate this calculus in an example. 
Consider the solution found in \cite{Argyres:2022lah} corresponding to $\D_{u,v}=\{4,5\}$,
\beq\label{crv45}
y^2=(v-u\, x)(x^5+u\, x-v) .
\eeq
The $x$ discriminant of the right hand side is
\beq\label{Dis45}
D_x:={\rm Disc}_x \big[(v-u\, x)(x^5+u\, x-v)\big] = v^{10} (256 u^5+ 3125 v^4),
\eeq
from which we determine that the CB described by this solution has two connected complex co-dimension one components, an unknotted and a knotted one. Furthermore $D_x$ vanishes with order ten at $v=0$ and order one at $256 u^5+ 3125 v^4=0$. 
  
The next step is to compare this with table \ref{tab:Kodaira} which lists the order of vanishing of the discriminant of the rank 1 CBs and specifies the linking monodromy. 
We then find that the knotted component is an $I_1$ singularity (the only one compatible with an order one vanishing) while the $v=0$ component can either be an $I_{10}$, an $I_4^*$ or a $II^*$. 
Since we have the explicit form of the curve, we can explicitly compute the monodromy at $v=0$ by following how the vanishing cycles transform as we take a small loop around $v=0$. 
After a short calculation we establish that the monodromy around $v=0$ is in fact an $I_{10}$. 

We now need to convert this monodromy information into a list of allowed deformations. This is a somewhat involved process which we have only briefly reviewed at the end of the previous subsection. 
The interested reader can consult appendix \ref{appB} or the original literature for further details.  The result is
\beq\label{Defs45}
\{4,5\}:\ \left\{
    \begin{array}{l}
        256 u^5+ 3125 v^4=0:\qquad U(1)\  w/\ 1\ {\rm hyper}\ w/\ q\text{=}1 \\
        v=0:\quad\left\{
        \begin{array}{l}
        {\boldsymbol a}.\ U(1)\  w/\ 10\ {\rm hypers}\ w/\ q\text{=}1 \\
        {\boldsymbol b}.\ U(1)\  w/\ 6\ {\rm hypers}\ w/\ q\text{=}1\ {\rm and}\ 1\ w/\ q\text{=}2 \\
        {\boldsymbol c}.\ U(1)\  w/\ 2\ {\rm hypers}\ w/\ q\text{=}1\ {\rm and}\ 2\ w/\ q\text{=}2 \\
        {\boldsymbol d}.\ U(1)\  w/\ 1\ {\rm hypers}\ w/\ q\text{=}1\ {\rm and}\ 1\ w/\ q\text{=}3 
        \end{array}
        \right.
   \end{array}
\right.
\eeq
giving rise to four different interpretations of the curve \eqref{crv45}, labeled from ${\boldsymbol a}$ to ${\boldsymbol d}$ above. We will continue the calculation below after introducing the \emph{central charge formulae} which convert the deformation datum into SCFT data.

\subsection{SCFT data and physical deformations}

Once we list the allowed deformations for each given curve, we can nearly completely characterise the conformal data of the rank 2 theories with those deformations. 
By conformal data we mean the $(a,c)$ central charges as well as the global symmetry $\ff$ and the flavor level $k_\ff$.  This is done using the \emph{central charge formulae} \cite{Martone:2020nsy},
\begin{subequations}
\begin{align}
\label{actotaint}
24 a &= h -2 + 6 (\D_u+\D_v) +\sum_{i\in I}\D^{\rm sing}_i b_{\rm rank-1}^i  ,
\\\label{actotbint}
12 c &= 4 + h + \sum_{i\in I}\D^{\rm sing}_{i} b_{\rm rank-1}^i ,\\\label{actotcint}
k_\ff&=\sum_{i\in I_{\ff}}\frac{\D_i^{\rm sing}}{d_i\D_i} \left(k^i-T({\bf2}\bh_i)\right)+T({\bf2}\bh).
\end{align}
\end{subequations}
Here $h$ is the quaternionic dimension of the theory's extended Coulomb branch (ECB) and the sum is over the irreducible components of the discriminant $\cV_i$. To determine $h$ from the deformation is a bit tricky and we refer to \cite[appendix D]{Kaidi:2022sng} for a thorough discussion. 
Henceforth we will simply take
\beq\label{ECBr2}
h=\sum_{i\in I} h_i
\eeq
where $h_i$ are the ECB quaternionic dimensions of the individual rank-1 theories and $b_{\rm rank-1}^i$ is defined in \eqref{bir1}. Continuing, $\D^{\rm sing}_i$ is the scaling dimension of the polynomial describing the component, given in \eqref{Deltasing}, and all the remaining quantities indexed by $i$ refer to known quantities of rank 1 theories which describe the set of BPS states becoming massless along the given component of the discriminant locus and are tabulated in table \ref{tab:r1theories} for SCFTs and tables \ref{tab:U1IRFree} and \ref{tab:SU2IRFree} for rank 1 IR-free theories.

Not all the deformations $\{\cT^i_{\rm rank-1}\}$ compatible with the set of linking monodromies $\{\Msc_{SL(2,\Z)}^i\}$ give rise to physical SCFT data. 
Before formulating this extra physical condition, recall that the computation of the $(a,c)$ central charges is straightforward once the set of $\{\cT^i_{\rm rank-1}\}$ is given, but the determination of the exact flavor symmetry structure is more subtle. 
In fact, besides \eqref{actotcint}, the main tool at our disposal is the \emph{UV-IR simple flavor condition} \cite{Martone:2020nsy} which determines the global flavor symmetry in terms of the flavor symmetries of the various $\{\cT^i_{\rm rank-1}\}$. 
This last step is sensitive, for example, to whether the rank 1 theory supported on a given stratum is discretely gauged or not --- discrete gauging generally modifies the flavor symmetry of the theory \cite{Argyres:2016yzz}. 
Unfortunately at the moment we do not have good control of allowed discrete gaugings nor of the consistency conditions to apply in this case. 
For this reason, it would be preferable if our study did not rely on the detailed structure of the flavor symmetry and instead relied only on the knowledge of $(a,c)$. 
We will now describe how this can be achieved.
But since the following paragraphs are a bit technical, let's first give the non-technical gist of the argument. 

Given a deformation $\{\cT^i_{\rm rank-1}\}$ we have two independent ways to estimate bounds on $d_{{\rm HB}_{\rm SCFT}}$, the quaternionic dimension of the Higgs branch (HB) of the corresponding SCFT. 
A lower bound is given by the maximum dimension of the HB of the set of rank 1 theories, and an upper bound which is obtained using anomaly matching and the central charge formulae. 
Since the two are independent, they are not necessarily consistent, \emph{i.e.} it is possible that the lower bound is actually larger than the upper bound. 
In this case there is a contradiction and we discard the deformation as unphysical. 
Now let's make this sharper.

Using anomaly matching, it is possible to relate the value of $(a,c)$ of a given SCFT to the quaternionic dimension of the HB of said SCFT by
\beq\label{CCandHB}
24(c-a)_{\rm SCFT}=d_{{\rm HB}_{\rm SCFT}}+24(c-a)_{\varnothing}
\eeq
where the LHS refers to the $(a,c)$ central charges of the SCFT while in the RHS the subscript $\varnothing$ refers to the central charges of a possibly non-trivial $\cN{=}2$ SCFT $\cT_{\varnothing}$ supported on the generic point of the HB, \emph{i.e.} the points which are reached by fully Higgsing the theory. 
Henceforth we will drop the subscript SCFT, all unsubscripted quantities will refer to the rank 2 SCFT under examination. 
In many cases $\cT_{\varnothing}$ is trivial and the low-energy theory on a generic point of the HB is simply a set of free hypermultiplets in which case $24(c-a)$ gives precisely the quaternionic dimension of the theory. 
But in our case, we don't know what $\cT_\varnothing$ is.
But, happily, the possibilities are severely constrained.
Clearly the only allowed $\cT_{\varnothing}$ are those with trivial HBs otherwise there would still be allowed Higgsing directions, contradicting the assumption that the theory is supported on the generic point of the HB. 
Furthermore, since the rank cannot increase along the HB, $\cT_{\varnothing}$ should be at most rank 2. 
There are only six known candidates which match these two conditions:
\beq\label{RemnantTh}
\begin{array}{c|c|c|c}
\multicolumn{4}{c}{\Large {\rm Theory\ vs.\ } 24(c-a)} \\
\hline\hline
\multicolumn{2}{c|}{{\rm rank\ 1}} & \multicolumn{2}{c}{{\rm rank\ 2}}\\
\hline
\cT^{(1)}_{\varnothing,1}& \frac 15 &(A_1,A_4) & \frac 27\\[1mm]
\hline
1\ {\rm free\ vector}&\, \,-1\, \, &\ 2\ {\rm free\ vectors}&-2\\[1mm]
\hline
\, \, {\rm frozen}\ IV^*_{Q=1}\, \,  &-\frac 52&\, \, \USp(4)\, w/\, \frac12 {\bf 16}\, \, &\, \, -2\, \, \\
\hline\hline
\end{array}
\eeq
These theories are described either in appendix \ref{appB}, \cite{Argyres:rank1III} or \cite{Martone:2021r2}. Henceforth we will make the assumption that these are the only possible candidates for $\cT_{\varnothing}$. 

With this data at hand, we can readily convert the $(a,c)$ values computed from a given deformation to a set of possible HB dimensions of the corresponding theory 
\beq
\Big(\{\cT^i_{\rm rank-1}\},\cT_{\varnothing}\Big)\quad
\mapsto\quad
d_{\rm HB}[\cT_\varnothing],
\eeq
where we made it explicit that the RHS depends on the choice of the $\cT_\varnothing$ in \eqref{RemnantTh}.  
Notice that, since dim$_\H$HB$\in \N$, in most cases only a fraction of the theories in \eqref{RemnantTh} are even compatible. 
Among those which are, consider the maximum value
\beq\label{maxHB}
d_{\rm HB}^{\rm max} \doteq {\rm max}\big\{0,{\rm max}_{\cT_\varnothing}d_{\rm HB}[\cT_\varnothing]\big\}.
\eeq
where the outer maximum is taken to ensure that $d_{\rm HB}^{\rm max}\geq0$.
This defines the aforementioned upper bound on the HB dimension. 
It is obtained by picking the entry in \eqref{RemnantTh} with the most negative value of $24(c-a)$ compatible with a positive integer total HB dimension. 
In cases in which $24(c-a)\in \Z$, which are the overwhelming majority of the cases examined below, we have
\beq\label{maxInt}
d_{\rm HB}^{\rm max}={\rm max}\big\{0,24(c-a)+2\big\} .
\eeq

Now a given deformation $\{\cT^i_{\rm rank-1}\}$ also gives a lower bound for the dimension of the HB of the corresponding SCFT. 
By continuity, the HB of the SCFT should at least contain the HBs of the rank 1 theories supported on each singular stratum.
It then follows that a lower bound for the dimension of the HB of the SCFT is
\beq
d_{\rm HB}^{\rm min}\doteq{\rm max}_{\{\cT^i_{\rm rank-1}\}}d_{\rm HB}(\cT^i_{\rm rank-1}) ,
\eeq
where $d_{\rm HB}(\cT^i_{\rm rank-1})$ indicates the quaternionic dimension of the HB of the rank 1 theory supported on the $i$-th connected stratum $\cV_i$.

We can then compare this lower bound with \eqref{maxHB}.
If
\beq\label{consHB}
d_{\rm HB}^{\rm min} > d_{\rm HB}^{\rm max}
\eeq
we have a contradiction and we thus discard the deformation $\{\cT^i_{\rm rank-1}\}$.
A loophole in this reasoning is that we cannot exclude that there exist other currently unknown theories which need to be added to the possible assignments in \eqref{RemnantTh} with negative enough $24(c-a)$ to sufficiently increase \eqref{maxHB} and invert the sign in \eqref{consHB}. 
Although in certain cases this could be rigorously excluded by combining the $a$ theorem \cite{Komargodski:2011vj} and the Hofman-Maldacena bound \cite{Hofman:2008ar}, we have not performed this refined analysis and take the list \eqref{RemnantTh} as an assumption of our analysis.

In two instances corresponding to scaling dimensions $\{\frac 32,\frac52\}$ and $\{2,2\}$ we will discard deformations which pass \eqref{consHB} because they fail more subtle consistency conditions.  
We discuss these cases in appendix \ref{appA}.

\subsubsection{Continuing with $\D_{u,v}=\{4,5\}$}

After all this work, it is time to apply this machinery to the example.  In \eqref{Defs45} we have already listed the deformations compatible with the scale invariant curve in \eqref{crv45},
\beq\label{deforms45}
\begin{array}{c|c|c}
\#  & \cT_{\rm rank-1}^{\rm knot} &\cT_{\rm rank-1}^{v=0}\\
  \hline
 {\boldsymbol a}.    &U(1)\  w/\ 1\ \ q\text{=}1 & U(1)\  w/\ 10\ q\text{=}1 \\
 {\boldsymbol b}.    & U(1)\  w/\ 1\ \ q\text{=}1 & U(1)\  w/\ 6\ q\text{=}1\ \&\ 1\ q=2\\
 {\boldsymbol c}.    & U(1)\  w/\ 1\ \ q\text{=}1 & U(1)\  w/\ 2\ q\text{=}1\ \&\ 2\ q=2\\
 {\boldsymbol d}.    & U(1)\  w/\ 1\ \ q\text{=}1 & U(1)\  w/\ 1\ q\text{=}1\ \&\ 1\ q=3
\end{array}
\eeq
It is a useful exercise to apply the central charge formulae to the deformations above to derive the following SCFT data assignments compatible with the scale invariant curve \eqref{crv45},
\beq\label{CC45}
\{4,5\}:\quad
\begin{array}{c|c:c:c}
 \#&\ 12c\,\,\, &\ 24a\,\,\,  & \ h\,\,\, \\
 \hline
  {\boldsymbol a}. & 74 & 122 &0    \\
  {\boldsymbol b}. & 59 & 107 &0     \\
 {\boldsymbol c}.  & 44 & 92 &0     \\
  {\boldsymbol d}.  & 34 &  82 &0    \\
\end{array}
\eeq
where the $h$ is computed using \eqref{ECBr2}. Now we need to establish which of these entries is consistent with \eqref{consHB}. Since in \eqref{CC45} 12$c$ and 24$a$ are integers for all entries, we can apply \eqref{maxInt} to compute the upper bound in each case finding,
\beq\label{upper45}
\{4,5\}:\quad
\begin{array}{c|c}
 \#&\ d_{\rm HB}^{\rm max}\, \, \\
 \hline
  {\boldsymbol a}. & 28 \\
  {\boldsymbol b}. &  13   \\
 {\boldsymbol c}.  &   0      \\
  {\boldsymbol d}.  & 0    \\
\end{array}
\eeq
Since in this case all deformations involve Lagrangian theories, see \eqref{deforms45}, there are a variety of ways to compute the $d_{\rm HB}^{\rm min}$. 
Perhaps the most straightforward in this context is to use \eqref{CCandHB} along with the standard central charge assignments for vector and hypermultiplets,
\beq
\begin{array}{c|c|c}
{\rm type} & \ 12 c\, \, &\ 24 a \, \,\\
\hline
{\rm vector} & 2 & 5\\
{\rm hyper} & 1 & 1
\end{array}
\eeq
For all cases in \eqref{deforms45} we have that $\cT_\varnothing$ is trivial. 
$\cT^{\rm knot}_{\rm rank-1}$ has empty HB, thus the lower bound always comes from the quaternionic dimension of the HB of $\cT^{v=0}_{\rm rank-1}$ which is readily computed to be
\beq\label{lower45}
\{4,5\}:\quad
\begin{array}{c|c}
 \#&\ d_{\rm HB}^{\rm min}\, \, \\
 \hline
  {\boldsymbol a}. & 9 \\
  {\boldsymbol b}. &  6   \\
 {\boldsymbol c}.  &   3      \\
  {\boldsymbol d}.  & 1    \\
\end{array}
\eeq
It follows that in the last two cases \eqref{consHB} is not satisfied and thus these two deformations are discarded:
\beq\label{CC45}
\{4,5\}:\quad
\begin{array}{c|c:c:c}
 \#&\ 12c\,\,\, &\ 24a\,\,\,  & \ h\,\,\, \\
 \hline
  {\boldsymbol a}. & 74 & 122 &0    \\
  {\boldsymbol b}. & 59 & 107 &0     \\
\rcr {\boldsymbol c}.  & 44 & 92 &0     \\
\rcr  {\boldsymbol d}.  & 34 &  82 &0    \\
\end{array}
\eeq
So in what follows, entries shaded in red will be those which do not give a consistent interpretation once the full moduli space is considered since they would have a negative dimension HB. 

It is remarkable that the two allowed values precisely realise two existing SCFTs, ${\boldsymbol a}.$ realises the SCFT labeled as $\suf(10)_{10}$ in \cite{Martone:2021r2} and realised in class-S in \cite{Chacaltana:2010ks}, while ${\boldsymbol b}.$ realises a theory which can be engineered from 5d by compactifying with a $\Z_2$ twist \cite{Zafrir:2016wkk}.

\section{New theories}\label{sec:3}

In this section we discuss in detail the deformations of the solutions found in \cite{Argyres:2022lah} which lead to new rank 2 SCFTs. The remaining cases, which reproduce remarkably well the current knowledge of rank 2 $\cN{=}2$ SCFTs, are reported in appendix \ref{appA}. To follow this analysis the material reviewed in appendix \ref{appB} can be useful. To most readers, some of the arguments below might feel a touch magical; this is likely because we will only sketch the results. These techniques have now been used extensively, and are explained in more detail \emph{e.g.} in \cite{Martone:2021r2} or \cite{Kaidi:2021tgr}.

Before moving to an explicit analysis of each case, there are two general rules that we need to apply in each case:
\begin{itemize}
    \item[1.] {\bf CB after Higgsing.} Say we perform a partial Higgsing of a rank 2 SCFT with scaling dimensions $\D_{u,v}$ by turning on the Higgs moduli of a rank 1 theory supported on an unknotted singularity ($u=0$ or $v=0$).  Then the CB scaling dimensions of the theory we ``land'' on are constrained by the full moduli space structure.  They depend on whether the partial Higgsing in question is $a.$ rank-decreasing or $b.$ rank-preserving, \emph{i.e.}, contributes to the ECB.
    \begin{itemize}
        \item[$a.$] \emph{Rank-decreasing Higgsing:} Since the CB of the Higgsed theory lies over a $u=0$ ($v=0$) plane, its CB coordinate is identified with $v$ ($u$) and thus the CB scaling dimension of the rank 1 theory must be $\D_v$ ($\D_u$).
        
        \item[$b.$] \emph{Rank-preserving Higgsing:} In this case the situation is more subtle. The general case is that the CB of the original SCFT is an orbifold of the (rank 2) CB of the Higgsed theory.  See \cite{Giacomelli:2020jel} for a detailed discussion of this structure.  In our case we find that the non-trivial orbifold groups must be
        \begin{align}
            \tilde{I}_4\ &\to\ \Z_2 ,\nn\\
            I_1^*\ &\to\ \Z_3 ,\\
            IV^* &\to\ \Z_4 .\nn
        \end{align}
        The $\Z_n$ orbifold action on the CB of the Higgsed theory multiplies the CB scaling coordinates by phases proportional to their scaling dimensions, $u_j \mapsto \exp\{2\pi i\D_j\} u_j$.  Thus, for example, if say $\{\D_u,\D_v\}= \{4,5\}$, then the $\Z_2$ orbifold action maps $(u,v) \mapsto (u,-v)$, so the invariant coordinates on the CB of the original SCFT are $(\til u, \til v) = (u,v^2)$ so have dimensions $\{4,10\}$. 
    \end{itemize}
    
    \item[2.] {\bf Unhiggsable frozen singularities.} The degrees of freedom corresponding to undeformable singularities do not carry any flavor symmetry, thus they cannot be lifted either by mass deformation or spontaneous symmetry breaking.  For instance, all the theories in the $I_1^*$ series can be mass-deformed to give rise to an undeformable $I_1^*$ on the CB.  (The physical interpretation of the undeformable $I^*_1$ singularity will be recalled below.)  Thus there should be at least a partial Higgsing of rank 2 theories of such a series down to rank 1 theories of the same series. 
\end{itemize}

\subsection{$\D=\{4,10\}$}

For this pair of scaling dimensions we will find many new theories. The SK solution for this entries was found already in \cite{Argyres:2005x5} to be:
\beq\label{crv410}
y^2=x^5+(u\,x-v)^3
\eeq
This leads to the discriminant:
\beq\label{Dis410}
D_x:={\rm Disc}_x \big[x^5+(u\,x-v)^3\big] = v^{10}(108u^5+3125v^2)
\eeq
which indicates two irreducible components, an unknotted and a knotted one with order of vanishing respectively equal to ten and one. Analysing the curve further we find that the knotted component is an $I_1$ while the unknotted one is a $II^*$.  For the unknotted component this lifts the ambiguity of the possible interpretations --- an $I_{10}$, an $I_4^*$, or a the actual $II^*$ ---given by just analysing the order of vanishing of \eqref{Dis410}. 

We are now in a position to go through the possible rank 1 theories which can be supported on the irreducible components of the singular locus of \eqref{crv46}:
\beq\label{Defs410}
   \{4,10\}:\ \left\{
   \begin{array}{cl}
    4u^3+27v^2=0&\quad U(1)\  w/\ 1\ {\rm hyper}\ w/\ q\text{=}1 \\
    v=0&\left\{
    \begin{array}{l}
    {\boldsymbol a}.\ \cT^{(1)}_{E_8,1}\\
    {\boldsymbol b}.\ \cS^{(1)}_{E_6,2}\\
    {\boldsymbol c}.\ \cS^{(1)}_{D_4,3}\\
    {\boldsymbol d}.\ \cS^{(1)}_{A_2,4}
    \end{array}
    \right.
    \end{array}
    \right.
\eeq
where the nomenclature for rank 1 theories listed above is derived from \cite{Apruzzi:2020pmv,Giacomelli:2020jel}. The $\cT^{(1)}_{E_8,1}$ was first discussed in \cite{Minahan:1996cj}, the $\cS^{(1)}_{E_6,2}$ in \cite{Argyres:2007tq}, the $\cS^{(1)}_{D_4,3}$ in \cite{Chacaltana:2016shw,Argyres:rank1III}, and the $\cS^{(1)}_{A_2,4}$ in \cite{Argyres:rank1IV}. We can now compute the associated SCFT data for each of the four possibilities. Since we found that the central charges are integers, $d_{\rm HB}^{\rm max}$ can be again computed using \eqref{maxInt}.
To instead compute $d_{\rm HB}^{\rm min}$, we notice that as usual $\cT^{\rm knot}_{\rm rank-1}$ has trivial HB and does not contribute, while for $\cT^{\rm v=0}_{\rm rank-1}$ we can simply look at table \ref{tab:r1theories} to find
\beq
\left\{4,10\right\}:\quad
\begin{array}{c|c:c:c:c:c}
 \#&\ 12c\,\,\, &\ 24a\,\,\,  &\ h\,\,\, & \ d_{\rm HB}^{\rm max}\, \, & \ d_{\rm HB}^{\rm min}\, \,  \\
 \hline
  {\boldsymbol a}. & 124 & 202 &0 &48 & 29 \\
\rcg  {\boldsymbol b}. & 99 & 177 & 5 &23 &  16 \\
\rcg  {\boldsymbol c}. & 88 & 166 & 4 &12 & 9 \\
\rcg  {\boldsymbol d}. & 82 & 160 & 3 &6 & 4
\end{array}
\eeq
All of these deformations are consistent with \eqref{consHB} but only ${\boldsymbol a}.$ matches the SCFT data of a known rank 2 theory: the $D_1^{20}(E_8)$ theory \cite{Giacomelli:2017ckh}. An $E_7$ class-S realization is presented in \cite{Chacaltana:2017boe} and a realisation in type \emph{II}B string theory, where only a subset of the SCFT data was computed, can be found in \cite{Cecotti:2012jx,Cecotti:2013lda}. We now turn to analysing the remaining three theories.

\subsubsection{Theory 1}

\begin{figure}[t!]
\ffigbox{
\begin{subfloatrow}
\ffigbox[7cm][]{
\begin{tikzpicture}[decoration={markings,
mark=at position .5 with {\arrow{>}}}]
\begin{scope}[scale=1.5]
\node[bbc,scale=.5] (p0a) at (0,0) {};
\node[scale=.5] (p0b) at (0,-2) {};
\node[scale=.8] (t0b) at (0,-2.1) {$\cT_{\{4,10\}_1}$};
\node[scale=.8] (p1) at (-.7,-1) {$[I_1,\varnothing]$\ \ };
\node[scale=.8] (p2) at (.7,-1) {\ \ $\cS^{(1)}_{E_6,2}$};
\node[scale=.8] (t1c) at (-1,-1.7) {{\scriptsize$\big[u^5+v^2=0\big]$}};
\node[scale=.8] (t1c) at (.7,-1.7) {{\scriptsize$\big[v=0\big]$}};
\draw[red] (p0a) -- (p1);
\draw[red] (p0a) -- (p2);
\draw[red] (p1) -- (p0b);
\draw[red] (p2) -- (p0b);
\end{scope}
\begin{scope}[scale=1.5,xshift=2.5cm]
\node[scale=.5] (p0a) at (0,2) {};
\node[scale=.5] (p0b) at (0,-2) {};
\node[scale=.8] (t0a) at (0,2.1) {$\H^{\rm d_{HB}}$};
\node[scale=.8] (p0aa) at (0,1) {$\cT^{(1)}_{D_4,1}$};
\node[scale=.8] (t0b) at (0,-2.1) {$\cT_{\{4,10\}_1}$};
\node[scale=.8] (p1) at (0,0) {$\cS^{(1)}_{D_4,2}$};
\node[scale=.8] (p2) at (0,-1) {$\hat{\cT}_{E_6,2}$};
\node[scale=.8] (t0c) at (.3,1.5) {$\df_4$};
\node[scale=.8] (t1c) at (.3,.5) {$\cf_3$};
\node[scale=.8] (t2c) at (.3,-0.5) {$\ff_4$};
\node[scale=.8] (t1c) at (.7,-1.5) {$\red{\H^5/\Z_2\equiv\cf_5}$};
\draw[blue] (p0a) -- (p0aa);
\draw[blue] (p0aa) -- (p1);
\draw[blue] (p1) -- (p2);
\draw[blue] (p2) -- (p0b);
\end{scope}
\end{tikzpicture}}
{\caption{\label{CB4101}The Coulomb and Higgs stratification of $\cT_{\{4,10\}_1}$.}}
\end{subfloatrow}\hspace{1cm}%
\begin{subfloatrow}
\capbtabbox[7cm]{%
  \renewcommand{\arraystretch}{1.1}
  \begin{tabular}{|c|c|} 
  \hline
  \multicolumn{2}{|c|}{$\cT_{\{4,10\}_1}$}\\
  \hline\hline
  $(\D_u,\D_v)$  &\quad $\left(4,10\right)$\quad{} \\
  $24a$ &  177\\  
  $12c$ & 99 \\
$\ff_k$ & $\red{\spf(10)}_{11}$  \\ 
$d_{\rm HB}$& 21\\
$h$&5\\
$T({\bf2}\bh)$&1\\
\hline\hline
\end{tabular}
}{%
  \caption{\label{Cc4101}Central charges, CB parameters and ECB dimension.}%
}
\end{subfloatrow}}{\caption{\label{4101}Information about the $\cT_{\{4,10\}_1}$ theory.}}
\end{figure}
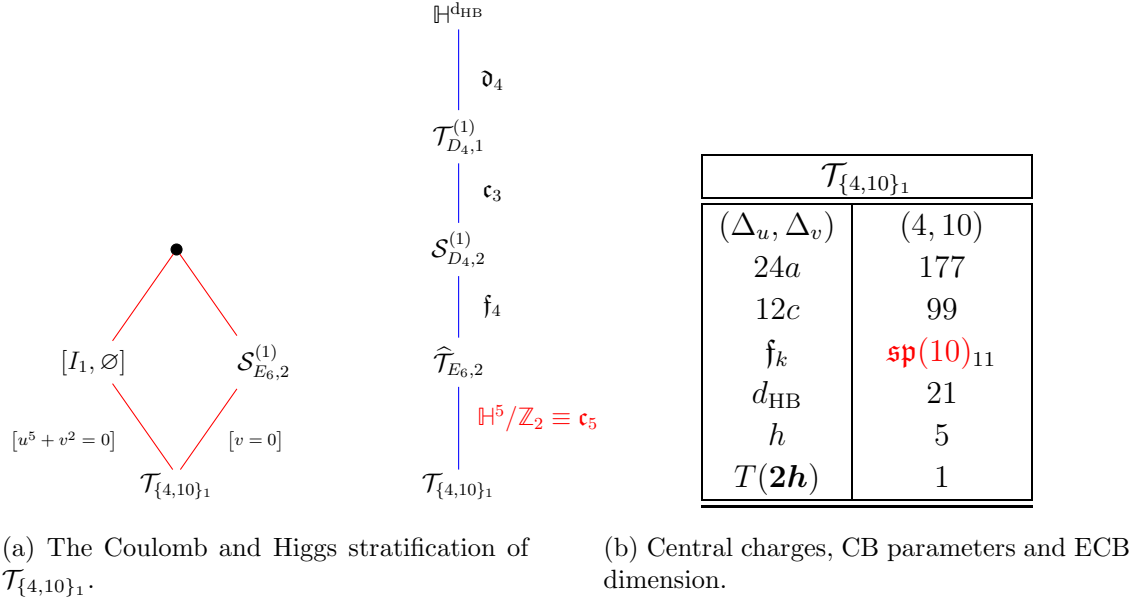

In order to describe this theory we will have to apply the rules described above. In this case, since we have a non-trivial ECB, we need to start with rule $1.b$. To describe how this works, introduce $\tilde{u}$ and $\tilde{v}$ to parametrise the CB coordinates of the rank-2 theory on which we land after the first partial Higgsing. Following \emph{e.g.} \cite{Giacomelli:2020jel}, we claim that the $\Z_2$ only acts on the $v$ and the $(\tilde{u},\tilde{v})$ are related to the original coordinates $(u,v)$: 
\beq\label{OrbECB}
u=\tilde{u}\quad{\rm and}\quad v=\tilde{v}^2
\eeq
we thus expect that the rank-2 theory supported on the ECB of $\cT_{\{4,10\}_1}$ has CB scaling dimension $(4,5)$. Using now rule $2.$ we can uniquely identify a candidate to be $\hat{\cT}_{E_6,2}$ since this theory higgses to $\cS^{(1)}_{D_4,2}$ which belongs to the $I_4$ rank-1 series, see \emph{e.g.} appendix \ref{appB}. This theory was first constructed in \cite{Chacaltana:2015bna,Wang:2018gvb} but this notation was introduced in \cite{Giacomelli:2020gee}. Its properties are:
\beq\label{proTE62}
\hat{\cT}_{E_6,2}:\quad \left\{
\begin{array}{cc}
(\D_u,\D_v) &(4,5)       \\
 24 a     & 112 \\
 12 c & 64 \\
 \ff_k & [\ff_4]_{10}\times\uf(1)\\
 d_{\rm HB} & 16\\
 h & 0
\end{array}
\right.
\eeq
For more detail on this theory we refer to the original papers or \cite{Martone:2021r2}.

Notice that once we identify the theory after the first partial Higgsing, the rest is tightly constrained. We can now determine that $\cT_\varnothing$ is trivial, that the dimension of the HB is 21 and in turn that the dimension of the first HB stratum is 5. Following the structure of the $\tilde{I}_4$ and of that the other series, we would expect this stratum to be $\H^5/\Z_2$. The case of $\Z_2$ is special, in fact $\H^n/\Z_2$ corresponds to the minimal nilpotent orbit of $\spf(2n)$, also labeled as $\cf_n$, making the first higgsing a \emph{generalised highest weight Higgsing} \cite{Martone:2021r2} which allows to use anomaly matching to further constrain the moduli space structure. But before reviewing how this works and applying it here, we need to determine the flavor symmetry of $\cT_{\{4,10\}_1}$. The level is straightforward and equal to 11, but to determine the form of the global symmetry we face an ambiguity related to possible discrete gaugings. By analysing the other theories in the $\tilde{I}_4$ series, see table \ref{tab:knownSWr2}, we find an intriguing pattern.

A naive matching with the structure of the various series --- see below --- would suggest a $\Z_2$ gauging of the rank-1 theory. In all cases of the $\tilde{I}_4$ series, the $\cT^{v=0}_{\rm rank-1}$ is an $\SU(2)$ $\cN=2$ gauge theory with only adjoint hypermultiplets for which the center symmetry does not act on local operators and can be gauged without affecting the flavor symmetry. We claim that the $\cS^{(1)}_{E_6,2}$ theory in the current setup, has also a $\Z_2$ which can be gauged without affecting the flavor symmetry, and thus determine the flavor symmetry of the theory to be $\spf(10)_{11}$ which also naturally acts on the first HB stratum.

We are now in a position of analysing the first higgsing and go back to \emph{generalised highest weight Higgsings} \cite{Martone:2021r2}. As mentioned, in these cases we can use anomaly matching to derive a simple formula determining the $c$ central charge of the rank 1 theory supported over the HB first symplectic stratum \cite{Giacomelli:2020jel}:
\beq\label{Higgs}
12 c_{\cT_{\rm Higgs}}=12c_{\cT_0}-2\left(\frac32 k_\ff -1\right)-{\rm dim}_\H \bar{\Sf} + 1
\eeq
where $\cT_0$ is the theory before Higgsing while $\cT_{\rm Higgs}$ is the theory after Higgsing. $k_\ff$ is the level of the simple flavor factor which is spontaneously broken along the Higgsing, \emph{i.e.}, $\spf(10)$ in this case, and ${\rm dim}_\H \bar{\Sf}$ is the quaternionic dimension of the symplectic stratum corresponding to the Higgsing.\footnote{The bar represents the set closure of $\Sf$ which is instead the symplectic leaf.  We will not delve into these details any longer; the interested reader can consult, for example \cite{Bourget:2019aer}.}

Applying \eqref{Higgs} to our case  $\cT_{\rm Higgs}\equiv \hat{\cT}_{E_6,2}$, $\cT_0\equiv \cT_{\{4,10\}_1}$, ${\rm dim}_\H \bar{\Sf}=5$ and $k_\ff=11$. We expect $\Sf$ to be the minimal nilpotent orbit of whatever flavor symmetry the theory has --- in this case the $\spf(10)$ minimal nilpotent orbit which is five quaternionic dimensional. Since we have already made a guess for the structure of the HB, \eqref{Higgs} acts as a consistency check, all in all we find that:
\beq\label{cc46}
64= 99 - 31 -4
\eeq
LHS and RHS match giving a beautifully consistent answer. To summarise things, the Hasse diagram of both the CB and HB is depicted in figure \ref{CB4101}.

\subsubsection{Theory 2}

\begin{figure}[t!]
\ffigbox{
\begin{subfloatrow}
\ffigbox[7cm][]{
\begin{tikzpicture}[decoration={markings,
mark=at position .5 with {\arrow{>}}}]
\begin{scope}[scale=1.5]
\node[bbc,scale=.5] (p0a) at (0,0) {};
\node[scale=.5] (p0b) at (0,-2) {};
\node[scale=.8] (t0b) at (0,-2.1) {$\cT_{\{4,10\}_2}$};
\node[scale=.8] (p1) at (-.7,-1) {$[I_1,\varnothing]$\ \ };
\node[scale=.8] (p2) at (.7,-1) {\ \ $\cS^{(1)}_{D_4,3}$};
\node[scale=.8] (t1c) at (-1,-1.7) {{\scriptsize$\big[u^5+v^2=0\big]$}};
\node[scale=.8] (t1c) at (.7,-1.7) {{\scriptsize$\big[v=0\big]$}};
\draw[red] (p0a) -- (p1);
\draw[red] (p0a) -- (p2);
\draw[red] (p1) -- (p0b);
\draw[red] (p2) -- (p0b);
\end{scope}
\begin{scope}[scale=1.5,xshift=2.5cm]
\node[scale=.5] (p0a) at (0,2) {};
\node[scale=.5] (p0b) at (0,-2) {};
\node[scale=.8] (t0a) at (0,2.1) {$\H^{\rm d_{HB}}$};
\node[scale=.8] (p0aa) at (0,1) {$\cT^{(1)}_{A_1,1}$};
\node[scale=.8] (t0b) at (0,-2.1) {$\cT_{\{4,10\}_2}$};
\node[scale=.8] (p1) at (0,0) {$\cS^{(1)}_{A_1,3}$};
\node[scale=.8] (p2) at (0,-1) {$\hat{\cT}_{D_4,3}$};
\node[scale=.8] (t0c) at (.3,1.5) {$\af_1$};
\node[scale=.8] (t1c) at (.5,.5) {$\H^2/\Z_3$};
\node[scale=.8] (t2c) at (.3,-0.5) {$\gf_2$};
\node[scale=.8] (t1c) at (.5,-1.5) {$\red{\H^4/\Z_3}$};
\draw[blue] (p0a) -- (p0aa);
\draw[blue] (p0aa) -- (p1);
\draw[blue] (p1) -- (p2);
\draw[blue] (p2) -- (p0b);
\end{scope}
\end{tikzpicture}}
{\caption{\label{CB4102}The Coulomb and Higgs stratification of $\cT_{\{4,10\}_2}$.}}
\end{subfloatrow}\hspace{1cm}%
\begin{subfloatrow}
\capbtabbox[7cm]{%
  \renewcommand{\arraystretch}{1.1}
  \begin{tabular}{|c|c|} 
  \hline
  \multicolumn{2}{|c|}{$\cT_{\{4,10\}_2}$}\\
  \hline\hline
  $(\D_u,\D_v)$  &\quad $\left(4,10\right)$\quad{} \\
  $24a$ &  166\\  
  $12c$ & 88 \\
$\ff_k$ & $\red{\suf(4)}_{11}$ \\ 
$d_{\rm HB}$& 10\\
$h$&4\\
$T({\bf2}\bh)$&1\\
\hline\hline
\end{tabular}
}{%
  \caption{\label{Cc4102}Central charges, CB parameters and ECB dimension.}%
}
\end{subfloatrow}}{\caption{\label{4102}Information about the $\cT_{\{4,10\}_2}$ theory.}}
\end{figure}
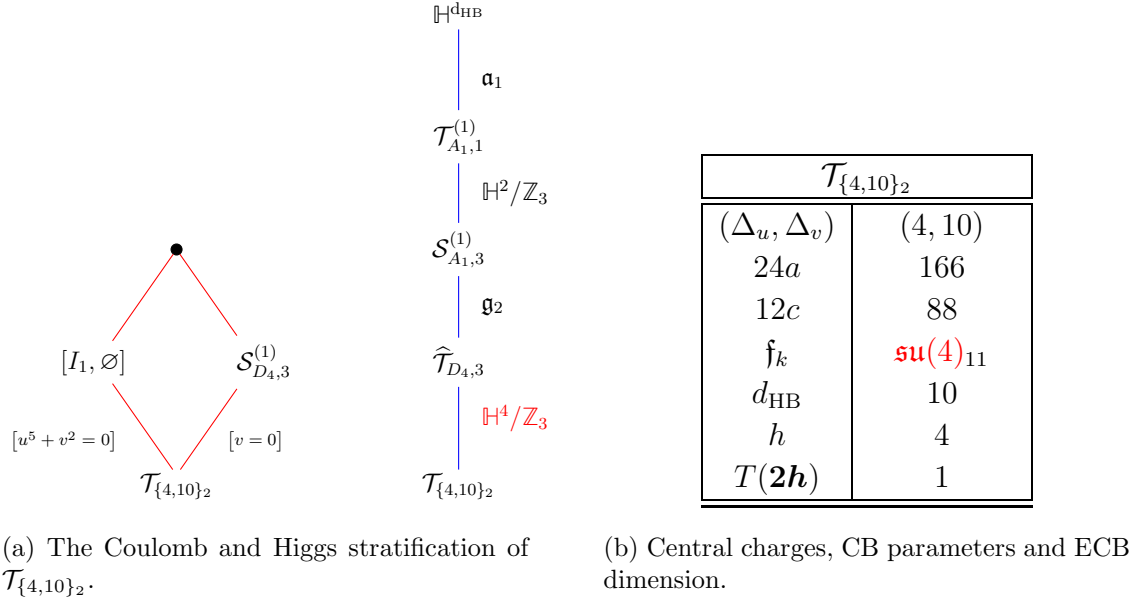

Now that we have also practised how to apply rule $1.b$ we can proceed more straightforwardly. An orbifold structure analogous to \eqref{OrbECB}, with $\Z_2\to \Z_3$, suggests that the rank-2 theory supported on the ECB of $\cT_{\{4,10\}_2}$ has CB scaling dimension $(4,\frac{10}3)$. Combining this observation with rule 2. we find that there is only one candidate at rank-2, see \cite{Martone:2021r2}, which matches just what we need: $\hat{\cT}_{D_4,3}$, found in \cite{Giacomelli:2020gee} and whose properties are:
\beq\label{proTD42}
\hat{\cT}_{D_4,3}:\quad \left\{
\begin{array}{cc}
(\D_u,\D_v) &(4,\frac{10}3)       \\
 24 a     & 82 \\
 12 c & 44 \\
 \ff_k & [\gf_2]_{\frac{20}3}\\
 d_{\rm HB} & 6\\
 h & 0
\end{array}
\right.
\eeq

This guess determines all the properties of the HB, including that the quaternionic dimension of the first stratum is four and which we guess to be $\H^4/\Z_3$. This would suggest that the flavor symmetry is $\suf(4)$ with level 11. Unfortunately this is not a generalised highest weight Higgsing so can't use \eqref{Higgs} to build an extra consistency check for our guess. By comparison with the other entries of the $I_1^*$ series --- briefly discussed below --- we would expect the rank-1 theory $\cT^{v=0}_{\rm rank-1}$ to be discretely gauged. Using the UV-IR simple flavor condition and the fact that we have determined $\ff$ to be $\suf(4)$, we claim that $\cS^{(1)}_{D_4,3}$ has a $\Z_3$ symmetry which can be gauged without affecting the flavor symmetry. This latter observation is quite speculative. The Hasse diagram of the CB and HB summarising our discussion is reported in figure \ref{CB4102}.

\subsubsection{Theory 3}

\begin{figure}[t!]
\ffigbox{
\begin{subfloatrow}
\ffigbox[7cm][]{
\begin{tikzpicture}[decoration={markings,
mark=at position .5 with {\arrow{>}}}]
\begin{scope}[scale=1.5]
\node[bbc,scale=.5] (p0a) at (0,0) {};
\node[scale=.5] (p0b) at (0,-2) {};
\node[scale=.8] (t0b) at (0,-2.1) {$\cT_{\{4,10\}_3}$};
\node[scale=.8] (p1) at (-.7,-1) {$[I_1,\varnothing]$\ \ };
\node[scale=.8] (p2) at (.7,-1) {\ \ $\cS^{(1)}_{A_2,4}$};
\node[scale=.8] (t1c) at (-1,-1.7) {{\scriptsize$\big[u^5+v^2=0\big]$}};
\node[scale=.8] (t1c) at (.7,-1.7) {{\scriptsize$\big[v=0\big]$}};
\draw[red] (p0a) -- (p1);
\draw[red] (p0a) -- (p2);
\draw[red] (p1) -- (p0b);
\draw[red] (p2) -- (p0b);
\end{scope}
\begin{scope}[scale=1.5,xshift=2.5cm]
\node[scale=.5] (p0a) at (0,1) {};
\node[scale=.5] (p0b) at (0,-2) {};
\node[scale=.8] (t0a) at (0,1.1) {$\uf(1)$};
\node[scale=.8] (t0b) at (0,-2.1) {$\cT_{\{4,10\}_3}$};
\node[scale=.8] (p1) at (0,0) {$\green{\cS^{(1)}_{\varnothing,4}}$};
\node[scale=.8] (p2) at (0,-1) {$\hat{\cT}_{A_2,4}$};
\node[scale=.8] (t1c) at (.5,.5) {$\H/\Z_4$};
\node[scale=.8] (t2c) at (.3,-0.5) {$\af_1$};
\node[scale=.8] (t1c) at (.5,-1.5) {$\red{\H^3/\Z_4}$};
\draw[blue] (p0a) -- (p1);
\draw[blue] (p1) -- (p2);
\draw[blue] (p2) -- (p0b);
\end{scope}
\end{tikzpicture}}
{\caption{\label{CB4103}The Coulomb and Higgs stratification of $\cT_{\{4,10\}_3}$.}}
\end{subfloatrow}\hspace{1cm}%
\begin{subfloatrow}
\capbtabbox[7cm]{%
  \renewcommand{\arraystretch}{1.1}
  \begin{tabular}{|c|c|} 
  \hline
  \multicolumn{2}{|c|}{$\cT_{\{4,10\}_3}$}\\
  \hline\hline
  $(\D_u,\D_v)$  &\quad $\left(4,10\right)$\quad{} \\
  $24a$ &  160 \\  
  $12c$ & 82 \\
$\ff_k$ & $\red{\suf(3)}_{11}$  \\ 
$d_{\rm HB}$& 5\\
$h$&3\\
$T({\bf2}\bh)$&1\\
\hline\hline
\end{tabular}
}{%
  \caption{\label{Cc4103}Central charges, CB parameters and ECB dimension.}%
}
\end{subfloatrow}}{\caption{\label{4103}Information about the $\cT_{\{4,10\}_3}$ theory.}}
\end{figure}
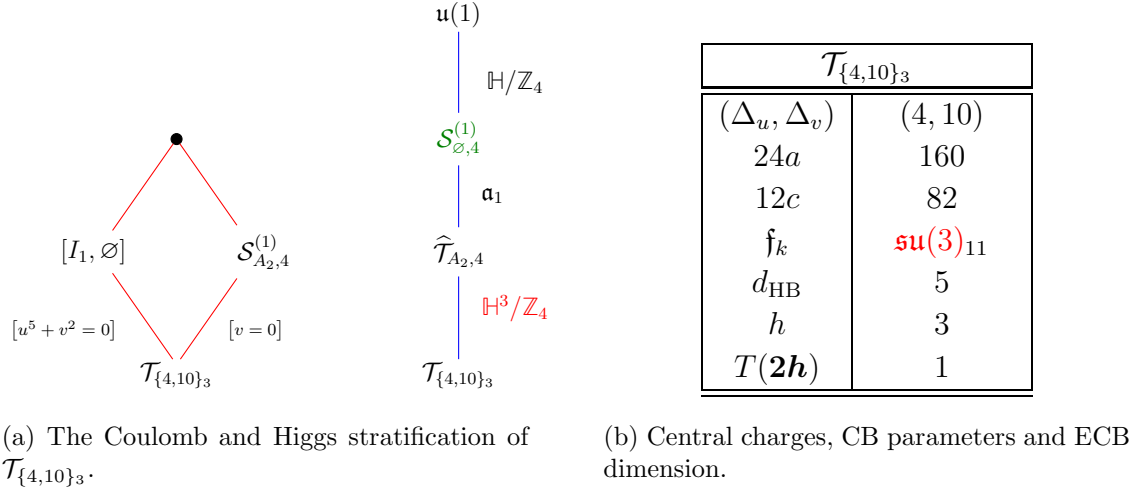

We have finally arrived at the analysis of the last deformation found in \eqref{Defs410}. This will comprise the sole --- for now --- element of the $IV^*$ rank-2 series. The discussion will be almost identical to the previous one.

Applying rules $1.b$ and $2.$, with an orbifold structure analogous to \eqref{OrbECB}, with $\Z_2\to \Z_4$, we identify $\hat{\cT}_{A_2,4}$ as the only guess which matches what we need, see \cite{Martone:2021r2} for details. This theory, found in \cite{Giacomelli:2020gee}, has properties:
\beq\label{proTD42}
\hat{\cT}_{D_4,3}:\quad \left\{
\begin{array}{cc}
(\D_u,\D_v) &(4,\frac{5}2)       \\
 24 a     & 67 \\
 12 c & 34 \\
 \ff_k & \suf(2)_5\\
 d_{\rm HB} & 2\\
 h & 0
\end{array}
\right.
\eeq
From the HB structure of $\hat{\cT}_{A_2,4}$, which can be found in \cite{Martone:2021r2}, we notice that in the case of $\cT_{\{4,10\}_3}$ $\cT_0$ is non-trivial, \emph{i.e.} a single vector multiplet. This is reflected in the calculation of the HB dimension, see table \ref{Cc4103}, as well as the Hasse diagram of its HB reported in figure \ref{CB4103}. The rest of the analysis follows what we already discussed in the previous section with $\Z_3\to \Z_4$ and $\suf(4)\to \suf(3)$. We summarise our results in figure \ref{CB4103}.

\subsection{New stratifications}

After discussing the prediction of new theories, we here analyse three more cases where our mass deformation approach enables us to sharpen our understanding of the stratification of rank-2 geometries.\footnote{We thank Craig Lawrie for comments prompting this realisation.}

\subsubsection{$\D=\{3,5\}$}

We will start by analysing a solution which appeared already in \cite{Argyres:2005x6},
\beq\label{crv35}
y^2=x\, \big(x^5 + (u\,x-v)^2 \big),
\eeq
with discriminant
\beq\label{Dis35}
D_x = v^9(108u^5+ 3125 v^3) ,
\eeq
which gives a singular locus with two irreducible components, an unknotted and a knotted one with order of vanishing respectively equal to nine and one. 
Matching the order of vanishing with possible rank 1 geometries, we find that the knotted component is an $I_1$, and after studying the behaviour of the non-trivial cycles to lift the ambiguity between an $I_9$, an $I_3^*$, and a $III^*$, we find that the unknotted one is an $I^*_3$. 

Now we run our algorithm to list the possible interpretations of this singularity data in terms of rank 1 theories, finding
\beq\label{Defs35}
    \left\{3,5\right\}:\quad\left\{
    \begin{array}{cl}
    108u^5+ 3125 v^3=0&\quad U(1)\  w/\ 1\ {\rm hyper}\ w/\ q\text{=}1 \\
    v=0&\left\{
    \begin{array}{l}
    {\boldsymbol a}.\ \quad SU(2)\  w/\ {\rm hyper}\ {\rm in\ the}\ 7\,{\bf 2}\\
    {\boldsymbol b}.\ \quad SU(2)\  w/\ {\rm hyper}\ {\rm in\ the}\ 3\,{\bf 2}\oplus{\bf 3}\\
    {\boldsymbol c}.\ \quad SU(2)\  w/\ {\rm hyper}\ {\rm in\ the}\ 4\,{\bf 3}\\
    {\boldsymbol d}.\ \quad SU(2)\  w/\ {\rm hyper}\ {\rm in\ the}\ 2\,{\bf 2}\oplus\tfrac{1}{2}\,{\bf 4}
   \end{array}
   \right.
   \end{array}
   \right.
   \eeq
Notice that it is possible to have a half hypermultiplet in the ${\bf 4}$ of $\SU(2)$ without introducing a Witten anomaly \cite{Witten:1982fp}.  In fact an $\SU(2)$ gauge theory with a single half-hyper in the ${\bf 4}$ has no $\cN{=}2$ preserving mass deformation and thus gives rise to a frozen $I_1^*$ singularity \cite{Argyres:rank1I}.  This leads to a whole series of rank 1 SCFTs --- the $I_1^*$ series \cite{Argyres:rank1III}.  Remarkably this structure will survive even in this more complicated context.
The associated SCFT data for each of the four possibilities is
\beq\label{defs35}
\left\{3,5\right\}:\quad
\begin{array}{c|c:c:c:c:c}
 \#&\ 12c\,\,\, &\ 24a\,\,\,  &\ h\,\,\, & \ d_{\rm HB}^{\rm max}\, \, & \ d_{\rm HB}^{\rm min}\, \, \\
 \hline
  {\boldsymbol a}. & 64 & 106 &0 &24 & 11 \\
  {\boldsymbol b}. & 50 & 92 & 1 &10 & 6 \\
  {\boldsymbol c}. & 53 & 95 & 4 &13 & 9 \\
  {\boldsymbol d}. & 44 & 86 & 0 &4 & 3
\end{array}
\eeq
These are all consistent with the HB dimension constraint \eqref{consHB}, and remarkably all match known SCFTs but we will see that ${\boldsymbol a}.$ matches the SCFT data of the $R_{2,5}$ theory \cite{Chacaltana:2015bna}, ${\boldsymbol b}.$ matches an SCFT which comes from compactification of a 5d SCFT \cite{Zafrir:2016wkk}, and ${\boldsymbol c}.$ matches the SCFT data of the $R_{2,4}$ theory introduced in \cite{Chacaltana:2014nya}. The last theory, ${\boldsymbol d}.$, matches the SCFT data of a theory found in \cite{Martone:20215D} but in this latter case, this bottom-up interpretation allows us to correct a mistake in stratification of the theory.

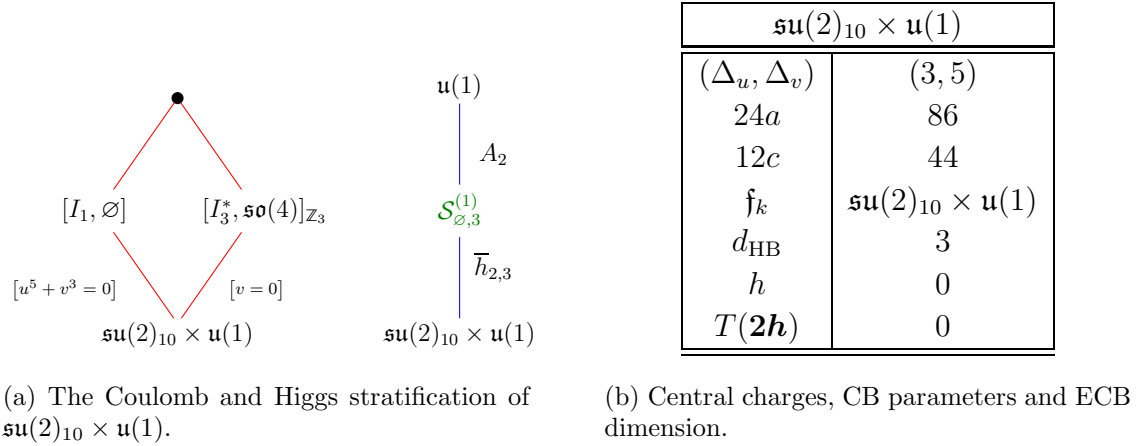
\begin{figure}[t!]
\ffigbox{
\begin{subfloatrow}
\ffigbox[7cm][]{
\begin{tikzpicture}[decoration={markings,
mark=at position .5 with {\arrow{>}}}]
\begin{scope}[scale=1.5]
\node[bbc,scale=.5] (p0a) at (0,0) {};
\node[scale=.5] (p0b) at (0,-2) {};
\node[scale=.8] (t0b) at (0,-2.1) {$\suf(2)_{10}\times\uf(1)$};
\node[scale=.8] (p1) at (-.7,-1) {$[I_1,\varnothing]$\ \ };
\node[scale=.8] (p2) at (.7,-1) {\ \ $[I_3^*,\sof(4)]_{\Z_3}$};
\node[scale=.8] (t1c) at (-1,-1.7) {{\scriptsize$\big[u^5+v^3=0\big]$}};
\node[scale=.8] (t1c) at (.7,-1.7) {{\scriptsize$\big[v=0\big]$}};
\draw[red] (p0a) -- (p1);
\draw[red] (p0a) -- (p2);
\draw[red] (p1) -- (p0b);
\draw[red] (p2) -- (p0b);
\end{scope}
\begin{scope}[scale=1.5,xshift=2.5cm]
\node[scale=.5] (p0a) at (0,0) {};
\node[scale=.5] (p0b) at (0,-2) {};
\node[scale=.8] (t0a) at (0,0.1) {$\uf(1)$};
\node[scale=.8] (t0b) at (0,-2.1) {$\suf(2)_{10}\times\uf(1)$};
\node[scale=.8] (p2) at (0,-1) {$\green{\cS^{(1)}_{\varnothing,3}}$};
\node[scale=.8] (t2c) at (.3,-0.5) {$A_2$};
\node[scale=.8] (t1c) at (.3,-1.5) {$\bar{h}_{2,3}$};
\draw[blue] (p0a) -- (p2);
\draw[blue] (p2) -- (p0b);
\end{scope}
\end{tikzpicture}}
{\caption{\label{CB35}The Coulomb and Higgs stratification of $\suf(2)_{10}\times\uf(1)$.}}
\end{subfloatrow}\hspace{1cm}%
\begin{subfloatrow}
\capbtabbox[7cm]{%
  \renewcommand{\arraystretch}{1.1}
  \begin{tabular}{|c|c|} 
  \hline
  \multicolumn{2}{|c|}{$\suf(2)_{10}\times\uf(1)$}\\
  \hline\hline
  $(\D_u,\D_v)$  &\quad $\left(3,5\right)$\quad{} \\
  $24a$ &  86\\  
  $12c$ & 44 \\
$\ff_k$ & $\suf(2)_{10}\times\uf(1)$ \\ 
$d_{\rm HB}$& 3\\
$h$&0\\
$T({\bf2}\bh)$&0\\
\hline\hline
\end{tabular}
}{%
  \caption{\label{Cc35}Central charges, CB parameters and ECB dimension.}%
}
\end{subfloatrow}}{\caption{\label{35}Information about the $\suf(2)_{10}\times\uf(1)$ theory. The stratification is new.}}
\end{figure}

The stratification for this theory was initially guessed in \cite{Martone:2021r2}. Since the matching of the central charges via the central charge formula only identified the $b$-factor \eqref{bir1} of the rank-1 theory supported on the $v=0$ singular locus, there is inherently an ambiguity in using this tool to infer the singular structure; this is what led to a wrong guess. In fact, in this case the initial guess was simply inconsistent as it seemed to lead to a non-scale invariant stratum. This inconsistency is resolved by the presence of the half hypermultiplet in the ${\bf 4}$. A similar mistake was made throughout the $I_1^*$ series as we will soon discuss. Notice that with this new interpretation, the order of the discriminant of the curve matches beautifully what is expected from the naive rank-1 analysis.

To obtain the flavor symmetry of the theory, $\suf(2)\times \uf(1)$, from the $\sof(4)$ flavor symmetry of the $\SU(2)$ with $2\,{\bf 2}\oplus\tfrac{1}{2}\,{\bf 4}$ gauge theory we expect a $\Z_3$ discrete gauging operation which is also consistent with the nature of the $I_1^*$ series. 
In fact, comparing with the results below, we find a very homogeneous behaviour where the $\sof(2n)$ flavor symmetry of the rank-1 $\SU(2)$ gauge theory with the half-hyper in the ${\bf 4}$ is discretely gauged to
\beq\label{DisSof}
\sof(2n)\quad\xrightarrow[]{\Z_3}\quad \uf(n) \ \text{or}\ 
 \suf(n) .
\eeq
The updated SCFT data about this theory $\{3,5\}$ is summarized in figure \ref{35}. It is straightforward to check that rules $1.a$ and $2.$ are satisfied by the stratification repored \ref{CB35}.

\subsubsection{$\D=\{4,6\}$}

Let's now turn to analysing a new solution found in \cite{Argyres:2022lah}:
\beq\label{crv46}
y^2=x\, (v-u\,x)\, (v-u\, x-x^3)
\eeq
which readily leads to the following discriminant:
\beq\label{Dis46}
D_x:={\rm Disc}_x \big[x\, (v-u\,x)\, (v-u\, x-x^3)\big] = v^{10}(4u^3+27v^2)
\eeq
this indicates two irreducible components, an unknotted and a knotted one with order of vanishing respectively equal to ten and one. Analysing the curve further we find that the knotted component is an $I_1$ while the unknotted is a $I_4^*$ which lift the ambiguity of the interpretation given just analysing the order of vanishing of \eqref{Dis46} between an $I_{10}$, an $I_4^*$ and a $II^*$.  Notice that the leading term in \eqref{crv46} vanishes as $u\to0$ and the curve degenerates to a genus 1 curve. Thus $u=0$ should be added as another component of the singular locus.  Upon a closer analysis we establish that this singularity is an $I_2$. 

We are now in a position to go through the possible rank 1 theories which can be supported on the irreducible components of the singular locus of \eqref{crv46}:
\beq\label{Defs46}
   \{4,6\}:\ \left\{
   \begin{array}{cl}
    4u^3+27v^2=0&\quad U(1)\  w/\ 1\ {\rm hyper}\ w/\ q\text{=}1 \\
    u=0&\quad U(1)\  w/\ 2\ {\rm hyper}\ w/\ q\text{=}1 \\
    v=0&\left\{
    \begin{array}{l}
    {\boldsymbol a}.\ \quad SU(2)\  w/\ {\rm hyper}\ {\rm in\ the}\ 8\,{\bf 2}\\
    {\boldsymbol b}.\ \quad SU(2)\  w/\ {\rm hyper}\ {\rm in\ the}\ 4\,{\bf 2}\oplus{\bf 3}\\
    {\boldsymbol c}.\ \quad SU(2)\  w/\ {\rm hyper}\ {\rm in\ the}\ 2\,{\bf 3}\\
    {\boldsymbol d}.\ \quad SU(2)\  w/\ {\rm hyper}\ {\rm in\ the}\ 5\,{\bf 3}\\
    {\boldsymbol e}.\ \quad SU(2)\  w/\ {\rm hyper}\ {\rm in\ the}\ 3\,{\bf 2}\oplus\tfrac{1}{2}\,{\bf 4}
    \end{array}
    \right.
    \end{array}
    \right.
\eeq
and we can now compute the associated SCFT data for each of the four possibility where, as usual, we can use \eqref{maxInt} to compute $d_{\rm HB}^{\rm max}$ and $\cT^{\rm knot}_{\rm rank-1}$ does not contribute to $d_{\rm HB}^{\rm min}$ since it has trivial HB:
\beq
\left\{4,6\right\}:\quad
\begin{array}{c|c:c:c:c:c}
 \#&\ 12c\,\,\, &\ 24a\,\,\,  &\ h\,\,\, & \ d_{\rm HB}^{\rm max}\, \, & \ d_{\rm HB}^{\rm min}\, \,  \\
 \hline
  {\boldsymbol a}. & 84 & 138 &0 &32 &  13 \\
  {\boldsymbol b}. & 67 & 121 & 1 &15 & 8 \\
\rcr  {\boldsymbol c}. & 50 & 104 & 2 &0& 3\\
  {\boldsymbol d}. & 71 & 125 & 5 &19& 12\\
  {\boldsymbol e}. & 60 & 114 & 0 &8 & 5
\end{array}
\eeq
Four out of these five deformations are consistent with \eqref{consHB}. Remarkably again all of the consistent theories can be matched with known SCFTs. Specifically ${\boldsymbol a}.$ matches the SCFT data of the theory labeled $\suf(2)_8\times \sof(16)_{12}$ in \cite{Martone:20215D}, for a class-S realisation see e.g. \cite{Chacaltana:2011ze}, ${\boldsymbol b}.$ with an SCFT labeled $\suf(4)_{12}\times \suf(2)_7\times \uf(1)$ in \cite{Martone:20215D}, for a class-S realisation see e.g. \cite{Chacaltana:2013oka} and ${\boldsymbol d}.$ with an SCFT labeled $\suf(2)_{8}\times \spf(10)_7$ in \cite{Martone:20215D}, for a class-S realisation see again \cite{Chacaltana:2013oka}. Finally ${\boldsymbol e}.$ matches the SCFT data of a theory first constructed in \cite{Zafrir:2016wkk} whose stratification was constructed in \cite{Martone:2021r2}. As in the previous case, the deformation in \eqref{Defs46} which corresponds with this entry, differs from the one initially inferred in \cite{Martone:2021r2}, although in this case the latter wasn't inherently inconsistent. 

\begin{figure}[t!]
\ffigbox{
\begin{subfloatrow}
\ffigbox[7cm][]{
\begin{tikzpicture}[decoration={markings,
mark=at position .5 with {\arrow{>}}}]
\begin{scope}[scale=1.5]
\node[bbc,scale=.5] (p0a) at (0,0) {};
\node[scale=.5] (p0b) at (0,-2) {};
\node[scale=.8] (t0b) at (0,-2.1) {$\suf(3)_{12}\times \uf(1)$};
\node[scale=.8] (p1) at (-1,-1) {$[I_1,\varnothing]$\ \ };
\node[scale=.8] (p2) at (1,-1) {\ \ $[I_4^*,\sof(6)]_{\Z_3}$};
\node[scale=.8] (p3) at (0,-1) {\ \ $[I_2,\suf(2)]$};
\node[scale=.8] (t1c) at (-1,-1.7) {{\scriptsize$\big[u^3+v^2=0\big]$}};
\node[scale=.8] (t1c) at (.8,-1.7) {{\scriptsize$\big[v=0\big]$}};
\node[scale=.8] (t1c) at (.27,-1.35) {{\scriptsize$\big[u=0\big]$}};
\draw[red] (p0a) -- (p1);
\draw[red] (p0a) -- (p2);
\draw[red] (p0a) -- (p3);
\draw[red] (p1) -- (p0b);
\draw[red] (p2) -- (p0b);
\draw[red] (p3) -- (p0b);
\end{scope}
\begin{scope}[scale=1.5,xshift=2.8cm]
\node[scale=.5] (p0a) at (0,1) {};
\node[scale=.5] (p0b) at (0,-2) {};
\node[scale=.8] (t0a) at (0,1.1) {$\H^{\rm d_{HB}}$};
\node[scale=.8] (t0b) at (0,-2.1) {$\suf(3)_{12}\times \uf(1)$};
\node[scale=.8] (p1a) at (-.7,0) {$\cT^{(1)}_{A_2,1}$};
\node[scale=.8] (p2a) at (-.7,-1) {$\cS^{(1)}_{A_2,3}$};
\node[scale=.8] (p1) at (.7,0) {$\cT^{(1)}_{A_1,1}$};
\node[scale=.8] (p2) at (.7,-1) {$\cS^{(1)}_{A_1,3}$};
\node[scale=.8] (t1c) at (.6,.5) {$\af_1$};
\node[scale=.8] (t1ca) at (-.6,.5) {$\af_2$};
\node[scale=.8] (t2c) at (1.2,-0.5) {$\H^2/\Z_3$};
\node[scale=.8] (t2ca) at (-1.2,-0.5) {$\H^3/\Z_4$};
\node[scale=.8] (t1c) at (.8,-1.5) {$\bar{h}_{3,3}$};
\node[scale=.8] (t1ca) at (-.6,-1.5) {$\af_1$};
\draw[blue] (p0a) -- (p1);
\draw[blue] (p0a) -- (p1a);
\draw[blue] (p1) -- (p2);
\draw[blue] (p1a) -- (p2a);
\draw[blue] (p2a) -- (p0b);
\draw[blue] (p2) -- (p0b);
\end{scope}
\end{tikzpicture}}
{\caption{\label{CB64}The Coulomb and Higgs stratification of $\suf(3)_{12}\times \uf(1)$.}}
\end{subfloatrow}\hspace{1cm}%
\begin{subfloatrow}
\capbtabbox[7cm]{%
  \renewcommand{\arraystretch}{1.1}
  \begin{tabular}{|c|c|} 
  \hline
  \multicolumn{2}{|c|}{$\suf(3)_{12}\times \uf(1)$}\\
  \hline\hline
  $(\D_u,\D_v)$  &\quad $\left(4,6\right)$\quad{} \\
  $24a$ &  114\\  
  $12c$ & 60 \\
$\ff_k$ & $\suf(3)_{12}\times \uf(1)$ \\ 
$d_{\rm HB}$& 6\\
$h$&0\\
$T({\bf2}\bh)$&0\\
\hline\hline
\end{tabular}
}{%
  \caption{\label{46Cc}Central charges, CB parameters and ECB dimension.}%
}
\end{subfloatrow}}{\caption{\label{46tot}Information about the $\suf(3)_{12}\times \uf(1)$ theory, the CB stratification is new.}}
\end{figure}
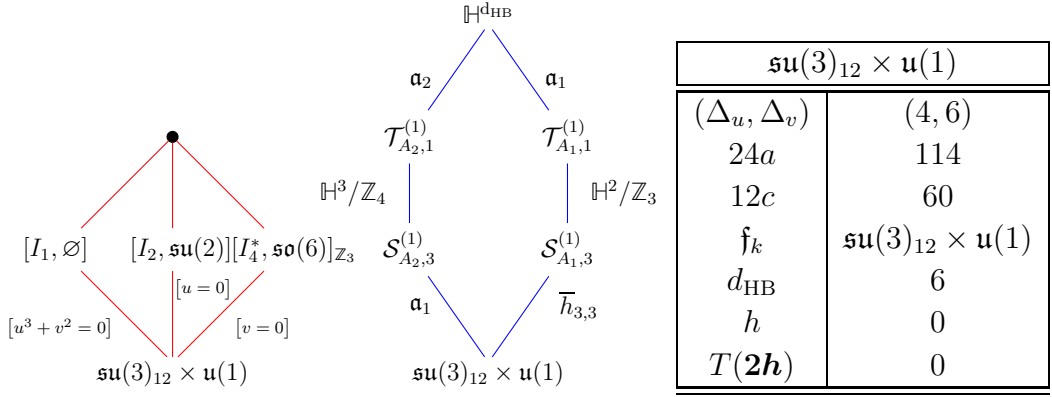

Even in this instance, in order to match the flavor symmetry of the theory, we need to postulate that the rank-1 theory is discretely gauged. We find that a pattern similar to the previous case works, 
\beq
\sof(6)\quad\xrightarrow[]{\Z_3}\quad \suf(3)\times\uf(1).
\eeq
The initial stratification guessed in \cite{Martone:2021r2} had a $I_0^*$ stratum rather than $I_4^*$ one. 
It is a straightforward calculation to show that interpreting the former as an $\SU(2)$ with $4\,{\bf2}$ gives the same $b_i$ \eqref{bir1} as interpreting the $I_4^*$ as an $\SU(2)$ with $3\,{\bf2}\oplus\frac12{\bf 4}$; this explains why the mistake was initially made. 
This new stratification provides a perfect matching between the order of vanishing of the discriminant of the rank-2 curve and the one which can be naively guessed by matching the rank-1 counting. 
Figure \ref{46tot} summarises the correct stratification, which is fully consistent with rules $1.a$ and $2.$ above, as well as the SCFT data of the theory.


\subsubsection{$\D=\{6,8\}$}

Let's now move to the last solution which leads to the prediction of a new SCFT,
\beq\label{crv68}
y^2=x\, (v-u\,x)\, (v-u\, x-x^4) .
\eeq
This is a solution which first appeared in \cite{Argyres:2022lah}. Taking the $x$ discriminant of the RHS leads to the following discriminant:
\beq\label{Dis68}
D_x:={\rm Disc}_x \big[x\, (v-u\,x)\, (v-u\, x-x^4)\big] = v^{12}(-27u^4+256v^3)
\eeq
which indicates two irreducible components, an unknotted and a knotted one with order of vanishing respectively equal to twelve and one. A more in-depth analysis of the curve reveals that the knotted component is an $I_1$ while the unknotted is a $I_6^*$ which lifts the ambiguity of the interpretation given by just analysing the order of vanishing of \eqref{Dis68} between an $I_{12}$ and an $I_6^*$.  We are now in a position to go through the possible rank 1 theories which can be supported on the irreducible components of the singular locus of \eqref{crv68},
\beq\label{Defs68}
   \{6,8\}:\ \left\{
   \begin{array}{cl}
    256v^3-27u^4=0&\quad U(1)\  w/\ 1\ {\rm hyper}\ w/\ q\text{=}1 \\
    v=0&\left\{
    \begin{array}{l}
    {\boldsymbol a}.\ \quad SU(2)\  w/\ {\rm hyper}\ {\rm in\ the}\ 10\,{\bf 2}\\
    {\boldsymbol b}.\ \quad SU(2)\  w/\ {\rm hyper}\ {\rm in\ the}\ 6\,{\bf 2}\oplus{\bf 3}\\
    {\boldsymbol c}.\ \quad SU(2)\  w/\ {\rm hyper}\ {\rm in\ the}\ 2\,{\bf 2}\oplus{\bf 3}\\
    {\boldsymbol d}.\ \quad SU(2)\  w/\ {\rm hyper}\ {\rm in\ the}\ 7\,{\bf 3}\\
    {\boldsymbol e}.\ \quad SU(2)\  w/\ {\rm hyper}\ {\rm in\ the}\ 5\,{\bf 2}\oplus\tfrac{1}{2}\,{\bf 4}\\
    {\boldsymbol f}.\ \quad SU(2)\  w/\ {\rm hyper}\ {\rm in\ the}\ 1\,{\bf 4}
    \end{array}
    \right.
    \end{array}
    \right.
\eeq
and we can now compute the associated SCFT data for each of the four possibilities as well as $d_{\rm HB}^{\rm max}$ and $d_{\rm HB}^{\rm mix}$ as we did in the previous cases:
\beq
\left\{6,8\right\}:\quad
\begin{array}{c|c:c:c:c:c}
 \#&\ 12c\,\,\, &\ 24a\,\,\,  &\ h\,\,\, & \ d_{\rm HB}^{\rm max}\, \, & \ d_{\rm HB}^{\rm min}\, \,  \\
 \hline
  {\boldsymbol a}. & 124 & 202 &0 &48 & 17 \\
  {\boldsymbol b}. & 101 & 179 & 1 &25  &12 \\
\rcr  {\boldsymbol c}. & 78 & 156 & 2 &2&4 \\
  {\boldsymbol d}. & 107 & 185 & 7 &31& 18  \\
  {\boldsymbol e}. & 92 & 170 & 0 &16& 9\\
\rcr {\boldsymbol f}. & 60 & 138 & 0 &0 & 1
\end{array}
\eeq
This gives a rich set of possibilities out of which only four are consistent with \eqref{consHB} and they can all be matched with known SCFTs. Specifically ${\boldsymbol a}.$ matches the SCFT data of the theory labeled $\sof(16)_{16}$ in \cite{Martone:20215D}, for a class-S realisation see e.g. \cite{Chacaltana:2014jba},  ${\boldsymbol b}.$ with an SCFT labeled $\suf(6)_{16}\times \suf(2)_9$ in \cite{Martone:20215D}, for alternative realisations see e.g. \cite{Chacaltana:2015bna,Ohmori:2018ona} and  ${\boldsymbol d}.$ with an SCFT labeled $\spf(14)_9$ in \cite{Martone:20215D}, and for alternative realisations see e.g. \cite{Ohmori:2018ona}. 
Finally ${\boldsymbol e}.$ matches instead a theory which was first constructed from 5d in \cite{Zafrir:2016wkk} and later characterised from 6d compactification in \cite{Ohmori:2018ona, Heckman:2022suy}. 
Again the stratification analysis made initially in \cite{Martone:2021r2} was wrong and it will be corrected here.

The first guess in \cite{Martone:2021r2} involved an $I_2^*$ where the low-energy theory was interpreted as an $\SU(2)$ with $6\,{\bf2}$. 
This is not inconsistent per-s\'e (as it was the initial guess for the $\suf(2)_{10}\times \uf(1)$) but our analysis here shows that it was mistaken; the correct guess being, from \eqref{Defs68}, a $I_6^*$ interpreted as an $\SU(2)$ with $5\,{\bf 2}\oplus \frac12{\bf 4}$. 
Not surprisingly the two have the same $b_i$ \eqref{bir1}.

\begin{figure}[t!]
\ffigbox{
\begin{subfloatrow}
\ffigbox[7cm][]{
\begin{tikzpicture}[decoration={markings,
mark=at position .5 with {\arrow{>}}}]
\begin{scope}[scale=1.5]
\node[bbc,scale=.5] (p0a) at (0,0) {};
\node[scale=.5] (p0b) at (0,-2) {};
\node[scale=.8] (t0b) at (0,-2.1) {$\suf(5)_{16}$};
\node[scale=.8] (p1) at (-.7,-1) {$[I_1,\varnothing]$\ \ };
\node[scale=.8] (p2) at (.7,-1) {\ \ $[I^*_6,\sof(10)]_{\Z_3}$};
\node[scale=.8] (t1c) at (-1,-1.7) {{\scriptsize$\big[u^4+v^3=0\big]$}};
\node[scale=.8] (t1c) at (.7,-1.7) {{\scriptsize$\big[v=0\big]$}};
\draw[red] (p0a) -- (p1);
\draw[red] (p0a) -- (p2);
\draw[red] (p1) -- (p0b);
\draw[red] (p2) -- (p0b);
\end{scope}
\begin{scope}[scale=1.5,xshift=2.5cm]
\node[scale=.5] (p0a) at (0,1) {};
\node[scale=.5] (p0b) at (0,-2) {};
\node[scale=.8] (t0a) at (0,1.1) {$\H^{\rm d_{HB}}$};
\node[scale=.8] (t0b) at (0,-2.1) {$\suf(5)_{16}$};
\node[scale=.8] (p1) at (0,0) {$\cT^{(1)}_{D_4,1}$};
\node[scale=.8] (p2) at (0,-1) {$\cS^{(1)}_{D_4,3}$};
\node[scale=.8] (t1c) at (.3,.5) {$\df_4$};
\node[scale=.8] (t2c) at (.5,-0.5) {$\H^4/\Z_3$};
\node[scale=.8] (t1c) at (.5,-1.5) {$\bar{h}_{5,3}$};
\draw[blue] (p0a) -- (p1);
\draw[blue] (p1) -- (p2);
\draw[blue] (p2) -- (p0b);
\end{scope}
\end{tikzpicture}}
{\caption{\label{CB68}The Coulomb and Higgs stratification of $\suf(5)_{16}$}}
\end{subfloatrow}\hspace{1cm}%
\begin{subfloatrow}
\capbtabbox[7cm]{%
  \renewcommand{\arraystretch}{1.1}
  \begin{tabular}{|c|c|} 
  \hline
  \multicolumn{2}{|c|}{$\suf(5)_{16}$}\\
  \hline\hline
  $(\D_u,\D_v)$  &\quad $\left(6,8\right)$\quad{} \\
  $24a$ & 170\\  
  $12c$ & 92 \\
$\ff_k$ & $\suf(5)_{16}$ \\ 
$d_{\rm HB}$& 14\\
$h$&0\\
$T({\bf2}\bh)$&0\\
\hline\hline
\end{tabular}
}{%
  \caption{\label{Cc68}Central charges, CB parameters and ECB dimension.}%
}
\end{subfloatrow}}{\caption{\label{68}Information about the $\suf(5)_{16}$ theory. The CB stratification is new.}}
\end{figure}
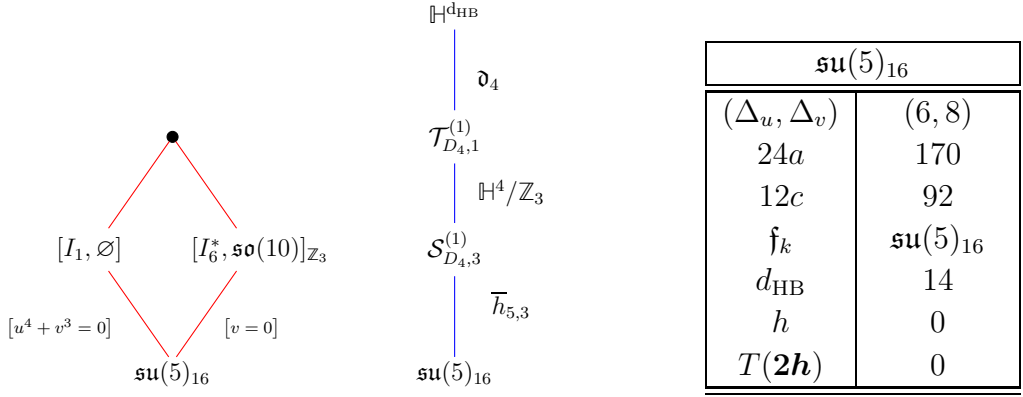

The two interpretations do differ in their global symmetries, and we guess that even in this case, as in the two previous instances of the $I_1^*$ series, the rank-1 theory supported on $v=0$ is actually discretely gauged as in \eqref{DisSof},
\beq
\sof(10)\quad \xrightarrow[]{\Z_3}\quad \suf(5) .
\eeq
We summarise the corrected stratification, which nicely passes both rule $1.a$ and rule $2.$, in figure \ref{68}.

\section{The landscape of rank 2 geometries}\label{sec5}

\begin{figure}[t!]
\centering
\begin{adjustbox}{center,max width=.9\textwidth}
\begin{tikzpicture}
[
auto,
good/.style={rectangle,rounded corners,fill=green!50,inner sep=2pt},
bad/.style={rectangle,rounded corners,fill=red!15,inner sep=2pt},
ugly/.style={rectangle,rounded corners,fill=blue!10,inner sep=2pt},
Marrow/.style={->,>=stealth[round],shorten >=1pt,line width=.4mm,coolblack},
IRarrow/.style={->,>=stealth[round],shorten >=1pt,line width=.4mm,deepcarmine,dashed}
]
\begin{scope}[yshift=.9cm,xshift=4 cm]
\node at (4.5,0) {{\large {\fontfamily{qcs}\selectfont\textsc{RG-flows for 4$d$  $\cN=2$ rank-2 SCFTs with $\varkappa=\{1,2\}$}}}};
\end{scope}


\begin{scope}[yshift=-.5cm,xshift=-2cm] 
\filldraw [fill=white, draw=black] (0.25,.2) rectangle (7,-7.25);
\node (E820) at (1,-1) [,align=center] {\scriptsize$[\ef_8]_{20}$};
\node (SO1410) at (3.5,-2) [,align=center] {\scriptsize$\sof(14)_{10}{\times}\uf(1)$};
\fill[color=yellow!40, rounded corners] (3,-2.7) rectangle (4,-3.3);
\node (U66) at (3.5,-3) [,align=center] {\scriptsize$\uf(6)_6$};
\node (D2A4) at (3.5,-4) [,align=center] {\scriptsize$D_2(\suf(5))$};
\node (A1D6) at (3.5,-5) [,align=center] {\scriptsize$(A_1,D_6)$};
\node (A1A5) at (4.25,-6) [,align=center] {\scriptsize$(A_1,A_5)$};
\node (A1A4) at (4.25,-7) [,align=center] {\scriptsize$(A_1,A_4)$};
\node (A1D5) at (2.75,-6) [,align=center] {\scriptsize$(A_1,D_5)$};
\fill[color=yellow!40, rounded corners] (.4,-2.7) rectangle (1.6,-3.3);
\node (SO128) at (1,-3) [,align=center] {\scriptsize$\sof(12)_8$};
\node (SO1612) at (3.5,-1) [,align=center] {\scriptsize$\sof(16)_{12}{\times}\suf(2)_8$};
\node (SO2016) at (4.25,0) [,align=center] {\scriptsize$\sof(20)_{16}$};
\node (SU1010) at (6,-1) [,align=center] {\scriptsize$\suf(10)_{10}$};
\node (SU88) at (6,-2) [,align=center] {\scriptsize$\suf(2)_6{\times}\suf(8)_8$};
\fill[color=yellow!40, rounded corners] (5.4,-2.7) rectangle (6.6,-3.3);
\node (SU245) at (6,-3) [,align=center] {\scriptsize$\suf(2)_4^5$};
\filldraw [fill=white, draw=black] (.25,.6) rectangle (7,.2);
\node[scale=.8] (title1) at (3.375,.4) {$I_1$\ series};
\end{scope}

\begin{scope}[yshift=-.5cm,xshift=.5cm] 
\filldraw [fill=white, draw=black] (5,.2) rectangle (11,-3.35);
\node (SU616) at (8,0) [,align=center] {\scriptsize$\suf(6)_{16}{\times}\suf(2)_9$};
\node (SU412) at (6.5,-1) [,align=center] {\scriptsize$\suf(4)_{12}{\times}\suf(2)_7{\times}\uf(1)$};
\node (SU3102) at (9.5,-1) [,align=center] {\scriptsize$\suf(3)_{10}{\times}\suf(3)_{10}{\times}\uf(1)$};
\node (SU310) at (6.5,-2) [,align=center] {\scriptsize$\suf(3)_{10}{\times}\suf(2)_6{\times}\uf(1)$};
\node (SU282) at (9.5,-2) [,align=center] {\scriptsize$\suf(2)_8{\times}\suf(2)_8{\times}\uf(1)^2$};
\fill[color=yellow!40, rounded corners] (7.2,-2.75) rectangle (8.8,-3.25);
\node (U12) at (8,-3) [,align=center] {\scriptsize$\uf(1){\times}\uf(1)$};
\filldraw [fill=white, draw=black] (5,.6) rectangle (11,.2);
\node[scale=.8] (title1) at (7.5,.4) {$I_4$\ series};
\end{scope}

\begin{scope}[yshift=-.5cm,xshift=7cm] 
\filldraw [fill=white, draw=black] (5.2,.2) rectangle (9.3,-3.35);
\node (SP149) at (8,0) [,align=center] {\scriptsize$\spf(14)_9$};
\node (SP107) at (8,-1) [,align=center] {\scriptsize$\spf(10)_7{\times}\suf(2)_8$};
\node (SP66) at (10.7,-1) [,align=center] {\scriptsize$\spf(8)_7{\times}\suf(2)_5$};
\node (SP86) at (8,-2) [,align=center] {\scriptsize$\spf(8)_6{\times} \uf(1)$};
\fill[color=deepcarmine!40, rounded corners] (5.4,-1.7) rectangle (6.6,-2.3);
\node (T4101) at (6,-2) [,align=center] {\scriptsize$\cT_{\{4,10\}_1}$};
\fill[color=yellow!40, rounded corners] (6.5,-2.75) rectangle (7.5,-3.25);
\node (SP65) at (7,-3) [,align=center] {\scriptsize$\spf(6)_5$};
\filldraw [fill=white, draw=black] (5.2,.7) rectangle (9.3,.2);
\node[scale=.8] (title1) at (7.25,.45) {$\tilde{I}_4$\ series};
\end{scope}

\begin{scope}[yshift=-5.5cm,xshift=1cm] 
\filldraw [fill=white, draw=black] (5.6,.2) rectangle (9,-2.25);
\node (SU510) at (8,0) [,align=center] {\scriptsize$\suf(5)_{16}$};
\node (SU312) at (8,-1) [,align=center] {\scriptsize$\suf(3)_{12}{\times}\uf(1)$};
\node (SU210U1) at (8,-2) [,align=center] {\scriptsize$\suf(2)_{10}{\times} \uf(1)$};
\fill[color=deepcarmine!40, rounded corners] (5.7,-.7) rectangle (6.9,-1.3);
\node (T4102) at (6.3,-1) [,align=center] {\scriptsize$\cT_{\{4,10\}_2}$};
\filldraw [fill=white, draw=black] (5.6,.65) rectangle (9,.2);
\node[scale=.8] (title1) at (7.3,.4) {$I^*_1$\ series};
\end{scope}

\begin{scope}[yshift=-5.5cm,xshift=8.5cm]
\filldraw [fill=white, draw=black] (5.7,-.6) rectangle (7.3,-1.4);
\fill[color=deepcarmine!40, rounded corners] (5.9,-0.7) rectangle (7.1,-1.3);
\node (T4102) at (6.5,-1) [,align=center] {\scriptsize$\cT_{\{4,10\}_3}$};
\filldraw [fill=white, draw=black] (5.7,.-.15) rectangle (7.3,-.6);
\node[scale=.8] (title1) at (6.5,-.35) {$IV^*$\ series};
\end{scope}

\begin{scope}[yshift=-10.2cm,xshift=1cm] 
\node (G28) at (8,0) [,align=center] {\scriptsize$[\gf_2]_8{\times}\suf(2)_{14}$};
\node (G2203) at (8,-1) [,align=center] {\scriptsize$[\gf_2]_{\frac{20}3}$};
\end{scope}

\begin{scope}[yshift=-10.2cm,xshift=-6cm] 
\node (SP128) at (8,0) [,align=center] {\scriptsize$\spf(12)_8$};
\node (SP86p) at (8,-1) [,align=center] {\scriptsize$\spf(8)_6{\times}\suf(2)_8$};
\node (SP65p) at (8,-2) [,align=center] {\scriptsize$\spf(6)_5{\times}\uf(1)$};
\fill[color=yellow!40, rounded corners] (8.5,-2.7) rectangle (9.5,-3.3);
\node (SP44) at (9,-3) [,align=center] {\scriptsize$\spf(4)_4$};
\node (SP88) at (10.5,0) [,align=center] {\scriptsize$\spf(4)_7{\times}\spf(8)_8$};
\node (SP66p) at (10.5,-1) [,align=center] {\scriptsize$\suf(2)_5{\times}\spf(6)_6{\times}\uf(1)$};
\node (F410) at (10.5,-2) [,align=center] {\scriptsize$[\ff_4]_{10}{\times}\uf(1)$};
\end{scope}

\begin{scope}[yshift=-10.2cm,xshift=6cm] 
\node (SP1211) at (8,0) [,align=center] {\scriptsize$\spf(12)_{11}$};
\fill[color=yellow!40, rounded corners] (8.4,-.7) rectangle (9.6,-1.3);
\node (SP87) at (9,-1) [,align=center] {\scriptsize$\spf(8)_7$};
\fill[color=yellow!40, rounded corners] (6,-.7) rectangle (8,-1.3);
\node (SP4133) at (9,-2) [,align=center] {\scriptsize$\spf(4)_{\frac{13}3}$};
\node (SU272) at (9,-3) [,align=center] {\scriptsize$\suf(2)_{\frac72}\times \uf(1)$};
\node (SP45) at (7,-1) [,align=center] {\scriptsize$\spf(4)_5{\times}\sof(4)_8$};
\end{scope}

\node (remThs) at (8.5,-9) [,align=center] { \textsc{Remaining series}};
\draw[dashed] (.5,-9.3) to (16.5,-9.3);

\draw[Marrow] (SP87) to (SP4133);
\draw[Marrow] (SP4133) to (SU272);

\draw[IRarrow] (E820) to (SO128);
\draw[Marrow] (SO2016) to (SO1612);
\draw[Marrow] (SO2016) to (SU1010);
\draw[Marrow] (SO1612) to (SO1410);
\draw[Marrow] (SU1010) to (SU88);
\draw[Marrow] (SO1612) to (SU88);
\draw[IRarrow] (SO1410) to (U66);
\draw[IRarrow] (SO1410) to (SO128);
\draw[IRarrow] (SU88) to (SU245);
\draw[IRarrow] (SU88) to (U66);
\draw[Marrow] (U66) to (D2A4);
\draw[Marrow] (D2A4) to (A1D6);
\draw[Marrow] (A1D6) to (A1D5);
\draw[Marrow] (A1D6) to (A1A5);
\draw[Marrow] (A1A5) to (A1A4);
\draw[Marrow] (SU616) to (SU3102);
\draw[Marrow] (SU616) to (SU412);
\draw[Marrow] (SU412) to (SU310);
\draw[Marrow] (SU412) to (SU282);
\draw[Marrow] (SU3102) to (SU282);
\draw[IRarrow] (SU282) to (U12);
\draw[IRarrow] (SU310) to (U12);
\draw[Marrow] (SP149) to (SP107);
\draw[Marrow] (SP107) to (SP86);
\draw[Marrow] (SP66) to (SP86);
\draw[IRarrow] (SP86) to (SP65);

\draw[Marrow] (SU510) to (SU312);
\draw[Marrow] (SU312) to (SU210U1);
\draw[Marrow] (G28) to (G2203);
\draw[IRarrow] (SP1211) to (SP45);
\draw[IRarrow] (SP1211) to (SP87);
\draw[Marrow] (SP128) to (SP86p);
\draw[Marrow] (SP86p) to (SP65p);
\draw[IRarrow] (SP65p) to (SP44);
\draw[Marrow] (SP88) to (SP86p);
\draw[Marrow] (SP88) to (SP66p);
\draw[Marrow] (SP66p) to (SP65p);
\draw[IRarrow] (F410) to (SP44);
\draw[IRarrow] (T4101) to (SP65);


\begin{scope}[yshift=-13.7cm,xshift=-3.5cm]
\node (L1) at (5,-.75){};
\node (L2) at (6.1,-.75){};
\node (L3) at (5,-1.25){};
\node (L4) at (6.1,-1.25){};
\node[anchor=west] (Text2) at (6,-.75)  {\small{\textsc{: mass deformations involving two or more relevant parameters,}}};
\node[anchor=west] (Text3) at (6,-1.25) {\small{\textsc{: mass deformation among SCFTs.}}};
\draw[IRarrow] (L1) to (L2);
\draw[Marrow] (L3) to (L4);
\end{scope}
\end{tikzpicture}
\caption{Graphical depiction of the RG-relations among four dimensional rank-2 $\cN=2$ SCFTs with $\varkappa=\{1,2\}$. We shade in \yellow{yellow} $\cN=2$ Lagrangian theories, in \green{green} $\cN=3$ theories and in \blue{blue} $\cN=4$ theories and in \red{red} the new theories. We have also omitted all the SCFTs which have no known mass deformation to any other $\cN=2$ SCFTs with $\varkappa=\{1,2\}$ with the exception of the new theories discussed here. The four series in the bottom part of the figure have no interpretation in terms of rank-1 undeformable singularities, while the five in upper part are interpreted as $I_1$ (top-left), $I_4$ (top-right), $\tilde{I}_4$ (middle-right), $I_1^*$ (middle-middle) and a lone $IV^*$ series.}
\label{MapR2}
\end{adjustbox}
\end{figure}

Recently there has been considerable progress in mapping out rank 2 $\cN{=}2$ SCFTs. 
In particular, thanks also to the study of their 5d origin \cite{Martone:20215D}, this space has been organised in different ``series'' which have the property that all theories within a series are connected by RG flows. 
In this paper we have made further progress in matching these notions to the rank 1 story and in particular to the rank 1 deformation patterns. 
A summary of the space of rank 2 theories is depicted in figure \ref{MapR2} which is an updated version of the one in \cite{Martone:20215D} but in which we have omitted all $\cN=2$ SCFTs with $\varkappa\neq\{1,2\}$ or those with no known RG trajectory connecting them to SCFTs with $\varkappa=\{1,2\}$. 
We shaded in red the new theories and which currently have neither a string theory nor a higher dimensional realization. 
It is a challenge to improve on this situation by finding other constructions which could fill in these gaps.

This situation recalls that of the rank 1 theories after a full classification had shown that many theories were missing from the string theory side. 
The recently introduced construction dubbed \emph{SW-fold} \cite{Heckman:2022suy}, a generalisation of the \emph{S-fold}, only partially reproduces the currently known rank-2 theories. 
Our analysis also suggests a more straightforward understanding of the structure which was uncovered using the notion of undeformable CB singularity \cite{Argyres:rank1I, Argyres:rank1IV} in terms of some sort of generalised orbifold action by cyclic groups
\beq\label{genOrb}
        \begin{array}{ccc}
            I_1 & \to & \Z_1\equiv\varnothing \\
            I_4/\,\tilde{I}_4 & \to& \Z_2 \\
            I_1^* & \to & \Z_3 \\
            IV^* &\to & \Z_4 \\
        \end{array}
\eeq
where now the cyclic group not only acts by discretely gauging theories supported on singular strata of the moduli space, but also determines the detailed structure of strata, in particular of the HB. 

\acknowledgments PCA is supported by DOE grant DE-SC0011784, MM is supported by STFC grant ST/T000759/1.

\appendix

\section{Analysis of deformations}\label{appA}

We here report the analysis of mass deformations of the solutions found in \cite{Argyres:2022lah} and not analysed in section \ref{sec:3}. These all reproduce known SCFTs.

\subsection{$\{\frac87,\frac{10}7\}$}

The curve, already found in \cite{Argyres:2005x5}, is
\beq\label{crv87107}
y^2=x^5+u x -v .
\eeq
The discriminant of the RHS is
\beq\label{Dis87107}
D_x = 256 u^5+ 3125 v^4,
\eeq
from which we determine that $D_x$ only has one irreducible component and the order of vanishing there equals one. 
Doing the matching between order of vanishing and possible rank 1 geometries, we find that the knotted component is an $I_1$ singularity,
\beq\label{Defs87107}
   \left\{\frac87,\frac{10}7\right\}:\quad\  {\boldsymbol a}. \quad
        256 u^5+ 3125 v^4=0\quad U(1)\  w/\ 1\ {\rm hyper}\ w/\ q\text{=}1,
\eeq
which leads to the following possible central charge assignment with $d_{\rm HB}^{\rm max}$ computed using $\cT_0\equiv (A_1,A_4)$ --- the only choice which gives an integer dimensional HB dimension --- and $d_{\rm HB}^{\rm min}$ being trvially zero:
\beq
\left\{\frac87,\frac{10}7\right\}:\quad
\begin{array}{c|c:c:c:c:c}
 \#&\ 12c\,\,\, &\ 24a\,\,\,  &\ h\,\,\, & \ d_{\rm HB}^{\rm max}\, \, & \ d_{\rm HB}^{\rm min}\, \, \\
 \hline
  {\boldsymbol a}. & \frac{68}7 & \frac{134}7 &0 &0 & 0 \\
\end{array}
\eeq
This is obviously consistent with \eqref{consHB} and this curve is then identified with the SW curve of a specific rank 2 Argyres-Douglas theory often labeled as $(A_1,A_4)$ \cite{Eguchi:1996vu, Eguchi:1996ds, Cecotti:2010fi, Xie:2012hs}.

\subsection{$\{\frac65,\frac85\}$}

Again the solution for this case was found in \cite{Argyres:2005x5} to be
\beq\label{curve6585}
y^2=x(x^4+u x -v),
\eeq
which has discriminant
\beq\label{Dis6585}
D_x = v^2\,(27 u^4+ 256 v^3) .
\eeq
$D_x$ has one unknotted and one knotted irreducible component with orders of vanishing two and one,respectively. Analysing the monodromies of the curve, we find that the knotted component is an $I_1$ singularity while the $v=0$ component is an $I_2$.
This leads to the assignment of rank 1 theories
\beq\label{Defs6585}
   \left\{\frac65,\frac85\right\}:\quad\  {\boldsymbol a}. \left\{
   \begin{array}{cl}
   27u^4+256v^3=0&\quad U(1)\  w/\ 1\ {\rm hyper}\ w/\ q\text{=}1 \\
   v=0&\quad U(1)\  w/\ 2\ {\rm hyper}\ w/\ q\text{=}1
   \end{array}
   \right.
   \eeq
which leads to the following possible central charge assignment with $d_{\rm HB}^{\rm max}$ computed using $\cT_0\equiv \cT^{(1)}_{\varnothing,1}$ --- the only choice which gives an integer dimensional HB dimension --- and $d_{\rm HB}^{\rm min}$ being equal to the dimension of $\cT^{v=0}_{\rm rank-1}$:
\beq
\left\{\frac65,\frac85 \right\}:\quad
\begin{array}{c|c:c:c:c:c}
 \#&\ 12c\,\,\, &\ 24a\,\,\,  &\ h\,\,\, & \ d_{\rm HB}^{\rm max}\, \, & \ d_{\rm HB}^{\rm min}\, \,  \\
 \hline
  {\boldsymbol a}. & 12 & \frac{114}5 &0 &1 & 1  \\
\end{array}
\eeq
These values are consistent with \eqref{consHB} and immediately identify \eqref{curve6585} as the SW curve of the $(A_1,D_5)$ Argyres-Douglas theory \cite{Eguchi:1996vu, Eguchi:1996ds, Cecotti:2010fi, Xie:2012hs}. 

\subsection{$\{\frac54,\frac32\}$}

Another known solution which appeared in \cite{Argyres:2005x6} is
\beq\label{crv5432}
y^2=x^6+u x -v ,
\eeq
whose discriminant,
\beq\label{Dis5432}
D_x = 3125 u^6+ 46656 v^5 ,
\eeq
implies a single irreducible component with order of vanishing equal to one. Doing the matching between order of vanishing and possible rank 1 geometries, we find that the knotted component is an $I_1$ singularity,
\beq\label{Defs5432}
   \left\{\frac54,\frac32\right\}:\quad
     {\boldsymbol a}.\quad    3125 u^6+ 46656 v^5=0\quad U(1)\  w/\ 1\ {\rm hyper}\ w/\ q\text{=}1 ,
   \eeq
from which we can compute the associated SCFT data with $d_{\rm HB}^{\rm max}$ computed using \eqref{maxInt} and $d_{\rm HB}^{\rm min}$ being trivially zero:
\beq
\left\{\frac54,\frac32\right\}:\quad
\begin{array}{c|c:c:c:c:c}
 \#&\ 12c\,\,\, &\ 24a\,\,\,  &\ h\,\,\, & \ d_{\rm HB}^{\rm max}\, \, & \ d_{\rm HB}^{\rm min}\, \,  \\
 \hline
  {\boldsymbol a}. & \frac{23}2 & 22 &0 &3 & 0    \\
\end{array}
\eeq
This data is consistent with \eqref{consHB} and uniquely singles out the $(A_1,A_5)$ Argyres-Douglas theory as the SCFT associated to this SISK geometry.

\subsection{$\{\frac{4}{3},\frac53\}$}

Continuing, consider the curve
\beq\label{crv4353}
y^2=x(x^5+u x -v) ,
\eeq
which appeared already in \cite{Argyres:2005x6}. 
Taking the $x$ discriminant of the RHS we find
\beq\label{Dis4353}
D_x = v^2\,(256 u^5 - 3125 v^4)
\eeq
which indicates that the singular locus has an unknotted and a knotted irreducible component with orders of vanishing two and one, respectively. 
Analysing the monodromies of the curve, we find that the knotted component is an $I_1$ singularity while the $v=0$ component is an $I_2$, leading to the following assignment of rank 1 theories:
\beq\label{Defs4353}
   \left\{\frac43,\frac53\right\}:\quad\  {\boldsymbol a}. \left\{
   \begin{array}{cl}
   256 u^5 - 3125 v^4=0&\quad U(1)\  w/\ 1\ {\rm hyper}\ w/\ q\text{=}1 \\
   v=0&\quad U(1)\  w/\ 2\ {\rm hyper}\ w/\ q\text{=}1
   \end{array}
   \right.
   \eeq
This gives the following SCFT data with $d_{\rm HB}^{\rm max}$ computed using \eqref{maxInt} and $d_{\rm HB}^{\rm min}$ being equal to the dimension of $\cT^{v=0}_{\rm rank-1}$:
\beq
\left\{\frac43,\frac53 \right\}:\quad
\begin{array}{c|c:c:c:c:c}
 \#&\ 12c\,\,\, &\ 24a\,\,\,  &\ h\,\,\, & \ d_{\rm HB}^{\rm max}\, \, & \ d_{\rm HB}^{\rm min}\, \, \\
 \hline
  {\boldsymbol a}. & 14 & 26 &0 &4 &1 \\
\end{array}
\eeq
This SCFT data is consistent with \eqref{consHB} and precisely matches that of the $(A_1,D_6)$ Argyres-Douglas theory.

\subsection{$\{\frac32,\frac52\}$}

In \cite{Argyres:2022lah} we find the curve for this case to be
\beq\label{crv3252}
y^2=x^5 + (u\, x- v)^2 .
\eeq
This solution is not new and was already listed in \cite{Argyres:2005x5}. It has discriminant
\beq\label{Dis3252}
D_x = v^5(3125 v^3- 108 u^5),
\eeq
giving two irreducible components, an unknotted and a knotted one, with orders of vanishing equal to five and one, respectively. 
Doing the matching between order of vanishing and possible rank 1 geometries, we find that the knotted component is an $I_1$ singularity while the unknotted an $I_5$. 
All possible interpretations of this datum in terms of rank 1 theories supported on the two irreducible components are
\beq\label{Defs3252}
    \left\{\frac32,\frac52\right\}:\quad\left\{
    \begin{array}{cl}
    3125 v^3+ 108 u^5=0&\quad U(1)\  w/\ 1\ {\rm hyper}\ w/\ q\text{=}1 \\
    v=0&\left\{
    \begin{array}{l}
    {\boldsymbol a}.\ \quad U(1)\  w/\ 5\ {\rm hyper}\ w/\ q\text{=}1\\
    {\boldsymbol b}.\ \quad U(1)\ w/\ 1\ {\rm hypers}\ w/\ q\text{=}1\ {\rm and}\ 1\ w/\ q\text{=}2
   \end{array}
   \right.
   \end{array}
   \right.
   \eeq

To establish which deformation is physical we need to compute $d_{\rm HB}^{\rm max/min}$ for each entry and check whether  \eqref{consHB} is satisfied. Again in this case we can compute the $d_{\rm HB}^{\rm min}$ as before by inputting a trivial $\cT_\varnothing$.  The result is
\beq\label{data3252}
\left\{\frac32,\frac52\right\}:\quad
\begin{array}{c|c:c:c:c:c}
 \#&\ 12c\,\,\, &\ 24a\,\,\,  &\ h\,\,\, &\ d_{\rm HB}^{\rm max}\, \, &\ d_{\rm HB}^{\rm min}\, \,  \\
 \hline
  {\boldsymbol a}. & 24 & 42 &0 &8 & 4  \\
\rcr  {\boldsymbol b}. & \frac{32}2 & \frac{69}2 &0 &1  & 1 \\
\end{array}
\eeq
both of which are compatible with \eqref{consHB} though the second is obtained by choosing $\cT_\varnothing \equiv {\rm frozen} \ IV^*_{Q=1}$ which has its CB scaling dimension equal to 3. 
This contradicts rule $1.a$ explained at the beginning of section \ref{sec:3}, and thus gives rise to an inconsistent choice. 
The first entry matches the SCFT data of the $D_2(SU(5))$ theory \cite{Eguchi:1996vu, Cecotti:2012jx, Cecotti:2013lda}.

\subsection{$\{2,2\}$}

Let's now analyse now a case in which there are scaling dimensions equal to two. These are associated with exactly marginal operators of the SCFT and thus we expect to find a 2-parameter family of solutions in this case. In this case, as already discussed in \cite{Argyres:2022lah}, we unexpectedly find a 4-parameter family of curves,
\beq\label{crv22}
y^2=(u\, x-v)(x^5 + \t_1 x^3 + \t_2 x^2 + \t_3 x + \t_4).
\eeq
This solution was not found in \cite{Argyres:2005x6, Argyres:2005x5} but a specialization of it (with only the two expected couplings) was already found in a previous work \cite{Argyres:1999fc} based on a string construction.  
From \eqref{crv22} we compute the discriminant
\beq\label{Dis22}
D_x \propto \prod_{i=1}^5(u+\a_i v)^2
\eeq
with $\a_i$ which are functions of the $\t_i$'s. 
This implies that we have five irreducible components with order of vanishing equal to two. 
Since the scaling dimension of $u$ and $v$ are equal in this case, there is no intrinsic notion of knotted vs.\ unknotted. 
By analysing the curve in more depth we lift the possible ambiguity between $I_2$ and a $II$, establishing that all five irreducible components are $I_2$'s. 
We are then left with the following possibilities
\beq\label{Defs22}
    \left\{2,2\right\}:\quad\left\{
    \begin{array}{cl}
    {\boldsymbol a.}\quad u+\a_i v=0,\ i=1,...,5\ &\quad U(1)\  w/\ 2\ {\rm hyper}\ w/\ q\text{=}1 \\
    {\boldsymbol b.}\quad u+\a_i v=0,\ i=1,...,5\ &\quad U(1)\  w/\ 1\ {\rm hyper}\ w/\ q\text{=}\sqrt{2} \\
   \end{array}
   \right.
   \eeq
This in turn leads to the following possible SCFT data assignments, 
\beq
\left\{2,2\right\}:\quad
\begin{array}{c|c:c:c:c:c}
 \#&\ 12c\,\,\, &\ 24a\,\,\,  &\ h\,\,\, & \ d_{\rm HB}^{\rm max}\, \, & \ d_{\rm HB}^{\rm min}\, \,  \\
 \hline
  {\boldsymbol a}. & 24 & 42 &0 &8 &1  \\
\rcr {\boldsymbol b}. & 14 & 32 & 0 &0 &0 \\
\end{array}
\eeq
where to compute $d_{\rm HB}^{\rm max}$ we used \eqref{maxInt} while $d_{\rm HB}^{\rm min}$ is determined by the dimension of the HB of the rank-1 theories supported on any of the connected component of the singular locus (they are all the same).
Surprisingly both entries are compatible with \eqref{consHB} while we expected only one solution, matching the data of the $\SU(2)\times SU(2)$ gauge theory with one bifundamental and two fundamentals per $\SU(2)$ factor.  But upon further analysis, we can rule out the second entry, which is for this reason shaded in red, by the following argument.

Given the fact that the curve \eqref{crv22} depends on continuous parameters (actually too many!) it implies that we can take a weak coupling limit and understand the deformations \eqref{Defs22} in that limit. This does not immediately rule out irrational values for the electric charge of the hypers if there is a change of normalisation of the electromagnetic coupling which allows to map this interpretation to the standard one, \emph{i.e.}, case ${\boldsymbol a}.$. This for example happens in rank-1 \cite{Argyres:rank1I}. Unfortunately no such normalization is possible in this case, and thus entry ${\boldsymbol b}.$ is incompatible with a consistent weak coupling limit and must be discarded.

\subsection{$\{2,3\}$}\label{sec:23}

We now proceed to another pair of scaling dimensions with integer entries. Because of the presence of a CB scaling dimension with dimension two we expect to find a one-parameter family of solutions.
Since the degrees of the $\SU(3)$ Casimirs are $(2,3)$, we expect to find a set of SCFT data compatible with the two inequivalent $\SU(3)$ $\cN{=}2$ SCFTs, namely $\SU(3)$ with hypermultiplets either in $6\,{\bf 3}$ or in ${\bf 3}\oplus {\bf 6}$. 
The scale invariant solution initially found in \cite{Argyres:2005x6} and reported in \cite{Argyres:2022lah} is
\beq\label{crv23}
y^2=x^6+x^3\, (u\,x-v)\t  + (u\, x-v)^2 .
\eeq
As in the previous case, the free parameter $\t$ should be identified with the holomorphic gauge coupling of the $\SU(3)$ gauge group. \eqref{crv23} has discriminant
\beq\label{Dis23}
D_x \propto v^6(u^3+ \a_1 v^2)(u^3+ \a_2 v^2)
\eeq
where $\a_1$ and $\a_2$ depend non-trivially on $\t$. This expression implies that there are three irreducible components, an unknotted and two knotted ones, with orders of vanishing equal to six, one, and one respectively. 
Doing the matching between order of vanishing and possible rank 1 geometries, we find that the two knotted components are $I_1$s while, after studying the behaviour of the non-trivial cycles to lift the ambiguity between an $I_6$ and $I_0^*$, we establish that the unknotted is an $I_6$:
\beq\label{Defs23}
    \left\{2,3\right\}:\quad\left\{
    \begin{array}{cl}
    u^3+ \a_1 v^2=0&\quad U(1)\  w/\ 1\ {\rm hyper}\ w/\ q\text{=}1 \\
    u^3+ \a_2 v^2=0&\quad U(1)\  w/\ 1\ {\rm hyper}\ w/\ q\text{=}1 \\
    v=0&\left\{
    \begin{array}{l}
    {\boldsymbol a}.\ \quad U(1)\  w/\ 6\ {\rm hyper}\ w/\ q\text{=}1\\
    {\boldsymbol b}.\ \quad U(1)\ w/\ 2\ {\rm hypers}\ w/\ q\text{=}1\ {\rm and}\ 1\ w/\ q\text{=}2
   \end{array}
   \right.
   \end{array}
   \right.
   \eeq
This in turn leads to the following two SCFT data with $d_{\rm HB}^{\rm max}$ computed using \eqref{maxInt} and $d_{\rm HB}^{\rm min}$ corresponding to the dimension of the HB of $\cT^{v=0}_{\rm rank-1}$ since the $\cT^{\rm knot}_{\rm rank-1}$ has a trivial HB:
\beq
\left\{2,3\right\}:\quad
\begin{array}{c|c:c:c:c:c}
 \#&\ 12c\,\,\, &\ 24a\,\,\,  &\ h\,\,\, & \ d_{\rm HB}^{\rm max}\, \, & \ d_{\rm HB}^{\rm min}\, \,  \\
 \hline
  {\boldsymbol a}. & 34 & 58 &0 &12 & 5 \\
 {\boldsymbol b}. & 25 & 49 & 0 &3 & 2  \\
\end{array}
\eeq
Both entries are consistent with \eqref{consHB} and match beautifully with the SCFT data of an $\SU(3)$ gauge theory with hypermultiplets in the $6\,{\bf 3}$ (${\boldsymbol a.}$) and ${\bf 3}\oplus {\bf 6}$ (${\boldsymbol b.}$) which are in fact the only $\cN{=}2$ $\SU(3)$ conformal gauge theories besides the $\cN{=}4$ superYang-Mills theory, whose SW curve is different \cite{Argyres:2022fwy}.  These solutions are present in \cite{Argyres:2005x6}, and were first discussed in \cite{Argyres:1995wt, Landsteiner:1998pb}.

\subsection{$\{2,4\}$}

Here is another pair of scaling dimensions with one entry equal to $2$, so we expect to find a one-parameter family of solutions where the free parameter $\t$ should be identified with the gauge coupling of the $\Sp(4)$ gauge algebra. 
The latter can be identified since the order of its Casimirs are precisely $2$ and $4$. 
The SW curve solution is
\beq\label{crv24}
y^2=x \big( x^4 + \t x^2 (u\,x-v) + (u\, x - v)^2\big),
\eeq
which already appeared in \cite{Argyres:2005x5}, and has discriminant
\beq\label{Dis24}
D_x \propto v^8(u^2+\a_1\, v)(u^2+\a_2\, v) ,
\eeq
where $\a_1$ and $\a_2$ depend non-trivially on $\t$. This expression implies that there are three irreducible components, an unknotted and two knotted ones, with orders of vanishing equal to eight, one, and one, respectively. 
Analysing the curve in more depth we find that the two knotted components are $I_1$'s while the unknotted is an $I_2^*$ lifting the ambiguity between an $I_{8}$, $I_2^*$ and a $IV^*$ from the sole knowledge of the order of vanishing of the discriminant. 
This singularity structure is compatible with following assignments of rank 1 IR-free theories:
\beq\label{Defs24}
    \left\{2,4\right\}:\quad\left\{
    \begin{array}{cl}
    u^3+ \a_1 v^2=0&\quad U(1)\  w/\ 1\ {\rm hyper}\ w/\ q\text{=}1 \\
    u^3+ \a_2 v^2=0&\quad U(1)\  w/\ 1\ {\rm hyper}\ w/\ q\text{=}1 \\
    v=0&\left\{
    \begin{array}{l}
    {\boldsymbol a}.\ \quad SU(2)\  w/\ {\rm hyper}\ {\rm in\ the}\ 6\,{\bf 2}\\
    {\boldsymbol b}.\ \quad SU(2)\  w/\ {\rm hyper}\ {\rm in\ the}\ 2\,{\bf 2}\oplus{\bf 3}\\
    {\boldsymbol c}.\ \quad SU(2)\  w/\ {\rm hyper}\ {\rm in\ the}\ 3\,{\bf 3}\\
    {\boldsymbol d}.\ \quad SU(2)\  w/\ {\rm hyper}\ {\rm in\ the}\ 1\,{\bf 2}\oplus\tfrac{1}{2}\,{\bf 4}
   \end{array}
   \right.
   \end{array}
   \right.
   \eeq
This in turn leads to the following four sets of SCFT data with $d_{\rm HB}^{\rm max}$ computed using \eqref{maxInt} and $d_{\rm HB}^{\rm min}$ corresponding to the dimension of the HB of $\cT^{v=0}_{\rm rank-1}$ since the $\cT^{\rm knot}_{\rm rank-1}$ has a trivial HB:
\beq
\left\{2,4\right\}:\quad
\begin{array}{c|c:c:c:c:c}
 \#&\ 12c\,\,\, &\ 24a\,\,\,  &\ h\,\,\, & \ d_{\rm HB}^{\rm max}\, \, & \ d_{\rm HB}^{\rm min}\, \,  \\
 \hline
 {\boldsymbol a}. & 44 & 74 & 0 &16 & 9 \\
\rcr {\boldsymbol b}. & 29 & 59 & 1 &1 & 4 \\
 {\boldsymbol c}. & 35 & 65 & 3 &7 & 6 \\
\rcr {\boldsymbol d}. & 28 & 58 & 0 &0 & 1  \\
\end{array}
\eeq
Two out of the four solutions are consistent with \eqref{consHB} and thus physical. 
Both of match the SCFT data of superconformal $\Sp(4)$ gauge theories. 
Specifically ${\boldsymbol a.}$ with $\Sp(4)$ with hypermultiplets in the $6\,{\bf 4}$ and ${\boldsymbol b.}$ with $\Sp(4)$ with $3\,{\bf 5}$ hypers. These are only two out of the five possible SCFTs with gauge algebra $\spf(4)$.  These solution were studied originally in \cite{Argyres:1995fw} and are also presented in \cite{Argyres:2005x6}.

\subsection{$\{3,4\}$}\label{sec:34}

Finally, let's analyse the consistent deformations of a new solution found in \cite{Argyres:2022lah}:
\beq\label{34sol}
y^2=(u\, x-v)(x^4 + u\, x-v) .
\eeq
Its discriminant is
\beq\label{Dis45}
D_x = v^8  (27 u^4+ 256 v^3) ,
\eeq
from which we determine that $D_x$ vanishes with order eight at $v=0$ and with order one at the knotted singularity $ 27 u^4+ 256 v^3 = 0$. 
Doing the matching between order of vanishing and possible rank 1 geometries, we find that the knotted component is an $I_1$ singularity (the only one compatible with an order one vanishing) while the $v=0$ can either be an $I_8$, an $I_2^*$ or a $IV^*$. 
Analysing the behaviour of the non-contractible cycles of the curve we establish that the $v=0$ monodromy is an $I_8$. 
Notice, however, that the leading coefficient of the curve is proportional to $u$.
Thus, even though it is not captured by the \eqref{Dis45}, the $u=0$ locus is also part of the singular locus. 
Again an analysis of the monodromy of the vanishing cycles on the curve identifies this as an $I_2$ singularity. 
All in all we are led to the following possible interpretations:
\beq\label{Defs34}
   \{3,4\}:\ \left\{
   \begin{array}{cl}
        256v^3+27u^4=0&\quad U(1)\  w/\ 1\ {\rm hyper}\ w/\ q\text{=}1 \\
        u=0&\quad U(1)\  w/\ 2\ {\rm hyper}\ w/\ q\text{=}1 \\
        v=0&\left\{
        \begin{array}{l}
         {\boldsymbol a}.\,U(1)\  w/\ 8\ {\rm hypers}\ w/\ q\text{=}1 \\
         {\boldsymbol b}.\, U(1)\  w/\ 4\ {\rm hypers}\ w/\ q\text{=}1\ {\rm and}\ 1\ w/\ q\text{=}2 \\
         {\boldsymbol c}.\,U(1)\  w/\ 2\ {\rm hypers}\ w/\ q\text{=}2
        \end{array}
        \right.
   \end{array}
   \right.
   \eeq
Now $d_{\rm HB}^{\rm max}$ can be straightforwardly computed using \eqref{maxInt} but $d_{\rm HB}^{\rm min}$ corresponds to the dimension of the HB of $\cT^{v=0}_{\rm rank-1}$ for case ${\boldsymbol a}.$ and ${\boldsymbol b}.$ and to the HB dimension of either $\cT^{v=0}_{\rm rank-1}$ or $\cT^{u=0}_{\rm rank-1}$ for case ${\boldsymbol c}.$. This leads to the following SCFT data,
\beq
\{3,4\}:\quad
\begin{array}{c|c:c:c:c:c}
 \#&\ 12c\,\,\, &\ 24a\,\,\,  &\ h\,\,\, & \ d_{\rm HB}^{\rm max}\, \, & \ d_{\rm HB}^{\rm min}\, \,  \\
 \hline
  {\boldsymbol a}. & 54 & 90 &0&20 & 7\\
  {\boldsymbol b}. & 42 & 78 &0 &8 & 4 \\
\rcr  {\boldsymbol c}.  &30 & 66&0&0& 1  \\
\end{array}
\eeq
with only two entries consistent with \eqref{consHB}, both of which can be matched with known SCFTs. Specifically ${\boldsymbol a}.$ matches the SCFT data of the $R_{0,4}$ theory \cite{Chacaltana:2015bna} and ${\boldsymbol b}.$ an SCFT which comes from compactification of a 5\emph{d} SCFT \cite{Zafrir:2016wkk}.

\section{Review of rank 1}\label{appB}

\subsection{Scale-invariant rank 1 CBs and genus 1 curves}

Assuming the CB $\simeq \C$ as a complex manifold, let $u$ be its complex coordinate.
Conformal invariance and unitarity imply that $u$ has definite scaling dimension $\D_u\ge1$, and that the CB has the holomorphic $\C^*$  homothety $u \mapsto \l^{\D_u} u$ for $\l\in\C^*$. 
The geometry of the CB is thus that of a flat cone with tip at $u=0$ and opening angle $2\pi/\D_u$.

Low energy $\cN{=}2$ supersymmetry on the CB implies it is special Kahler (SK), meaning there is a pair of special coordinates on the CB, $\s^A$, with 
\beq\label{formR1}
\bpm \s^1\\ \s^2 \epm =
u^{1/\D_u} \bpm \t\\ 1 \epm
\eeq
where $\t$ is a constant, and $u^{1/\D_u}$ is the (local) flat complex coordinate on the cone.  
Furthermore, the special coordinates' monodromies are in the $\SL(2,\Z)$ EM duality group.
Upon continuing around a closed path linking $u=0$ once counterclockwise, the special coordinates transform as
\beq
\bpm \s^1\\ \s^2 \epm
\ \xrightarrow{\ \ \g\ \ }\ 
e^{i2\pi/\D_u} \bpm \s^1\\ \s^2 \epm
\equiv\ \mathscr{M}_\g
\bpm \s^1\\ \s^2\epm ,
\qquad \mathscr{M}_\g\in \SL(2,\Z).
\eeq
Thus $e^{i2\pi/\D_u}$ has to be an eigenvalue of an $\SL(2,\Z)$  matrix. 
It is a simple exercise to show that the only such eigenvalues are
$e^{i n\pi/3}$ or $e^{i n\pi/2}$ for $n\in\Z$, and since $\D_u\ge1$ we can immediately read off the list of allowed values of $\D_u$, which also specifies the special geometry using \eqref{formR1}.  These coincide with the first 7 rows in table \ref{tab:Kodaira}. 
To define the SK geometry we also need a choice of a holomorphic one-form and here we have chosen it to be $\w = dx/y$ in all cases.  
Also, in table \ref{tab:Kodaira} we have taken the generators of $\SL(2,\Z)$ to be the standard elements $S = \bspm 0 & -1\\1&0\espm$, $T = \bspm 1 & 1\\ 0 & 1\espm$.

\begin{table}[tp]
\hspace*{-0.8cm}
\def\arraystretch{1}%
\begin{tabular}{c|c|c|c|c|c|c|c|c|c}
Series & Name & Kodaira & $12\, c$ & $\D_u$ & $d_{\mathrm{HB}}$ & $h_{\mathrm{ECB}}$ & $b$ & $\gf$ & $k_{\mathfrak{g}}$  \\
\hhline{==========}
&$\cT^{(1)}_{E_8,1}$ &$II^*$&
62&6& 29&0 & 10&$\ef_8$ &12 \\[1mm]
&$\cT^{(1)}_{E_7,1}$ &$III^*$&
38&4& 17&0 &9&$\ef_7$&8\\[1mm]
&$\cT^{(1)}_{E_6,1}$ &$IV^*$&
26&3& 11&0 &8&$\ef_6$&6\\[1mm]
$I_1$&$\cT^{(1)}_{D_4,1}$ &$I_0^*$&
14&2& 5&0&6&$\sof(8)$&4\\[1mm]
&$\cT^{(1)}_{A_2,1}$ &$IV$&
8&$\frac32$& 2&0&4&$\suf(3)$&3\\[1mm]
&$\cT^{(1)}_{A_1,1}$ &$III$&
6&$\frac43$& 1&0&3&$\suf(2)$&$\frac83$\\[1mm]
&$\cT^{(1)}_{\varnothing,1}$&$II$&
$\frac{22}5$&$\frac65$& 0&0&2&$\varnothing$& \\[1mm]
\cdashline{1-10}
\multirow{4}{*}[-3mm]{$I_4$}&$\cS^{(1)}_{E_6,2}$&$II^*$&
49&6& 16&5&7&$\spf(10)$&7\\[1mm]
&$\cS^{(1)}_{D_4,2}$&$III^*$&
29&4& 8&3&6&$\spf(6){\times}\suf(2)$&(5,8)\\[1mm]
&$\cS^{(1)}_{A_2,2}$&$IV^*$&
19&3& 4&2&5&$\spf(4){\times} \uf(1)$&$(4,\star)$\\[1mm]
&$\blue{\cS^{(1)}_{\varnothing,2}}$&$I_0^*$&
9&2& 1&1&3&$\suf(2)$&3\\[1mm]
\cdashline{1-10}
&$\cS^{(1)}_{D_4,3}$&$II^*$&
42&6& 9&4&6&$\suf(4){\rtimes} \Z_2$&14\\[1mm]
$I^*_1$&$\cS^{(1)}_{A_1,3}$&$III^*$&
24&4& 4 &3&5&$\!\!\suf(2){\times}\uf(1){\rtimes}\Z_2\!\!$&$(10,\star)$\\[1mm]
&$\green{\cS^{(1)}_{\varnothing,3}}$&$IV^*$&
15&3& 1&1&4&$\uf(1)$&$\star$ \\[1mm]
\cdashline{1-10}
&$\cS^{(1)}_{A_2,4}$&$II^*$&
38&6& 5&3&$\frac{11}2$&$\suf(3){\rtimes} \Z_2$&14\\[1mm]
$IV^*_1$&$\green{\cS^{(1)}_{\varnothing,4}}$&$III^*$&
21&4& 1&1&$\frac92$&$\uf(1){\rtimes} \Z_2$&$\star$\\[1mm]
&$IV^*_{Q=1}$ &$IV^*$& 
$\frac{25}2$ & 3 & 0&0 &$\frac72 $& $\varnothing$ &
\end{tabular}
\caption{\small The list of rank-one SCFTs which can describe the low-energy physics on codimension-one singular loci of the Coulomb branch. Additional information can be found in \cite{Argyres:2016yzz}. Theories in \green{green} are $\cN=3$, while the  theory in \blue{blue} is $\cN=4$. We are here neglecting the possible discretely gauged variants.
\label{tab:r1theories}}
\end{table}%

Each scale invariant geometry corresponds to multiple SCFTs.
The full analysis was performed in \cite{Argyres:rank1I, Argyres:rank1II, Argyres:rank1III, Argyres:rank1IV}, and table \ref{tab:r1theories} summarizes the results.

\subsection{Rank one IR-free theories}
\label{app:IRFree}

In order to account for IR-free theories, where $\t = i\infty$, we need to also consider the $\SL(2,\Z)$ elements which are conjugate to $T$ transformations. In this case no scale-invariant solution for the special coordinates $\s$ exists, and we must include the leading corrections to scaling, $\s(u) \approx u^{\b_0} \ln^{\b_1}(u)$ as $u\to0$.  
Solutions for $T^n$ monodromies are
\begin{align}
\s \approx u \bpm \frac{n}{2\pi i} \ln(u)\\ 1 \epm
\qquad \text{as}\qquad u \to 0,
\end{align}
corresponding to $\D_u \approx 1$, which are IR-free $\U(1)$ gauge theories.
For this solution the metric is $ds^2 \approx -\frac{n}{4\pi} \ln(u\bar u)\,du\, d\bar u$ which is positive-definite in the vicinity of $u=0$ only for $n>0$.  Thus the $T^n$ monodromies for $n \in \Z^+$ give sensible geometries, and account for the $I_n$ geometries in table \ref{tab:Kodaira}.  A similar story goes for the $-T^n$ monodromies, which correspond to IR-free $\SU(2)$ gauge theories with massless hypermultiplets and correspond to the $I_n^*$ entries in the table. 

The rest of this appendix, which is largely taken from \cite[Appendix C]{Kaidi:2022sng} and \cite[Section 4.2]{Argyres:rank1I}, discusses the detailed interpretation of these rank one IR-free geometries. Let us begin with the case of $\U(1)$ gauge theory. The most general matter content consists of $n_i$ hypermultiplets of charge $q_i$ for $i=1, \dots, N$. The beta function for this theory is given by $\beta\propto\sum_i q_i^2 n_i$. The relevant Kodaira fiber is then of type $I_{\sum_i q_i^2 n_i}$, and the corresponding flavor symmetry is  $\bigoplus_i \u(n_i)$. By nature of being a $\U(1)$ theory we have $\D_u = 1$, and it is easy to see that  $12c = 2 + \sum_i n_i$ and $h_{\mathrm{ECB}}=0$.  This is enough data to compute $b= \sum_i n_i$ and $k_i = 2$. A summary of the CFT data for $\U(1)$ IR-free theories is given in table \ref{tab:U1IRFree}.

Next consider IR-free $\SU(2)$ gauge theories. 
The most general matter content consists of $r_i$ hypermultiplets in real (orthogonal) representations ${\bf R}_i$, and $s_j$ hypermultiplets in symplectic (pseudoreal) representations ${\bf S}_j$.  
Let the number of distinct real representations that occur be $R$, and the number of symplectic ones be $S$.
Denote by $d_{\bf R}$ and $T_{\bf R}$ the dimension and Dynkin index of representation $\mathbf R$.
Note that half-integer $s_j$, corresponding to ``half-hypermultiplets", are allowed as long as $\sum_j s_j T_{\bf S_j} \in\Z$, so there is no Witten anomaly.
The CB scaling dimension is $\D_u = 2$, 
the beta-function is proportional to $b_0 = \sum_i r_i T_{\mathbf R_i} + \sum_j s_j T_{\bf S_j}- 4$,
so the relevant Kodaira fiber is of type $I^*_{b_0}$.
The flavor symmetry is $\gf=\bigoplus_{i=1}^R \sp(2r_i) \bigoplus_{j=1}^S \so(2s_j)$. 
Each $\sp(2r_i)$ flavor factor contributes a factor of $r_i$ to the dimension of the extended Coulomb branch, so $h_{\mathrm{ECB}} = \sum_{i=1}^R r_i$, and the central charge receives contributions from all matter fields $12 c = 6 + \sum_i r_i\, d_{\bf R_i} + \sum_j s_j\, d_{\bf S_j}$. 
From this information we may then compute the quantity $b$ as well as the flavor levels. 
A summary of the CFT data for $\SU(2)$ theories is reported in table \ref{tab:SU2IRFree}.

\begin{table}[t]
\begin{adjustbox}{center,max width=2\textwidth}
\renewcommand{\arraystretch}{1.2}
\hspace*{-0.1cm}\begin{tabular}{|c|c|c|c|c|}
\multicolumn{5}{c}{$\cN{=}2$ $\U(1)$ gauge theories w/ $n_i$ hypers of charge $q_i$}\\
\hline
\hline
\multicolumn{5}{|c|}{Kodaira type $I_n$ with $n= \sum_i n_i q_i^2$}\\
\hline
\hline
$12 c$& $\D_u$ &$h_{\mathrm{ECB}}$&$b$&$k_{\mathfrak{g}}$\\
\hline
$2 + \sum_i n_i$& 1&  0 &$ \sum_i n_i $&  $(2, \dots, 2)$\\
\hline
\end{tabular}
\caption{CFT data of IR-free $\U(1)$ gauge theories.}
\label{tab:U1IRFree}
\end{adjustbox}
\end{table}%

\begin{table}[ht]
\begin{center}
\begin{tabular}{|c|c|c|c|c|}
\multicolumn{5}{c}{$\cN{=}2$ $\SU(2)$ gauge theories w/ hypers in $\oplus_i^R\, r_i \, {\bf R}_i \, \oplus_j^S\, s_j\, {\bf S}_j$}\\
\hline
\hline
\multicolumn{5}{|c|}{Kodaira type $I^*_n$ with $n = \sum_i r_i T_{{\bf R}_i}{+}\sum_j s_j T_{{\bf S}_j} {-}4$}\\
\hline
\hline
$12 c$ &$\D_u$ &$h_{\mathrm{ECB}}$ &$b$ &$k_\gf$\\
\hline
\rule{0em}{7mm}
$6{+}\sum_i r_i d_{{\bf R}_i} {+} \sum_j s_j d_{{\bf S}_j}$ &
2 &  
$\sum_i r_i$ &
$2{+}\sum_i\frac{r_i (d_{{\bf R}_i}-1)}{2} {+} \sum_j \frac{s_j \,d_{{\bf S}_j}}{2}$ & 
$(\underbrace{3, \dots, 3}_{R\ {\rm times}} , \underbrace{4, \dots, 4}_{S\ {\rm times}})$\\
\hline
\end{tabular}
\caption{CFT data of IR-free $\SU(2)$ gauge theories.
$R$ and $S$ refer to real and symplectic representations.
$d_{\bullet}$ and $T_{\bullet}$ denote dimension and Dynkin index.}
\label{tab:SU2IRFree}
\end{center}
\end{table}%

Note that in the $\SU(2)$ case there is an additional subtlety: changing the normalization of the generators can lead to different monodromies. 
This happens if we rescale the generator $t_3$ of the $\U(1)\subset \SU(2)$ by a factor $a$ so that $t_3 \to a t_3$.
Then $\U(1)$ charges on the CB will be rescaled by the same factor, and the one-loop beta function coefficient (which is proportional to a sum of squares of charges) will be rescaled by $a^2$.  
Thus an $I_n^*$ singularity will correspond to such a charge-rescaled $\SU(2)$ theory if
\begin{align}\label{I*n-su2}
n = a^2 b_0,
\end{align}
and will have $\U(1)$ charges on the CB given by 
\begin{align}\label{u1su2norm}
Q=2 a t_3 .
\end{align}

But not every value of $a$ is allowed, because the set of allowed $\U(1)$ charges on the CB is constrained by the Dirac quantization condition to be (integer multiples of) square roots of positive integers, $Q^2 \in \N$.   This leads to the following possibilities:
\begin{align}\label{a2cases}
a^2 &\in \tfrac14\, \N & &\text{if all hypermultiplets are in orthogonal irreps,}
\nonumber\\
a^2 &\in \N & &\text{otherwise.}
\end{align} 
The first case is allowed since the orthogonal irreps are odd-dimensional, and so the eigenvalues of $t_3$ are integers.  Note that this is the case where the global form of the gauge group is allowed to be ${\rm SO}(3) \cong {\rm SU}(2)/\Z_2$.  In the second case in \eqref{a2cases} the global form of the gauge group must be ${\rm SU}(2)$.

\bibliographystyle{Auxiliary/JHEP}

\end{document}